\documentclass[12pt, a4paper]{article}
\pdfoutput=1
\usepackage[utf8]{inputenc} 
\usepackage[T1]{fontenc}
\usepackage{amsmath}
\usepackage{amssymb}
\usepackage{amsfonts}
\usepackage{bbm}
\usepackage{soul}
\usepackage{verbatim}
\usepackage{dsfont}
\usepackage{booktabs}
\usepackage{slashed}
\usepackage{bm}
\usepackage{subcaption}
\usepackage{mathtools}
\usepackage{epsfig}
\usepackage{color}
\usepackage[table]{xcolor}
\usepackage{graphicx}
\usepackage[]{caption}
\usepackage{float}
\usepackage{overpic}
\usepackage{enumerate}
\usepackage{hhline}
\usepackage{multirow}
\usepackage{xspace}
\usepackage{setspace}
\usepackage[title]{appendix}
\usepackage{soul}
\usepackage{tensor}
\usepackage{tablefootnote}
\usepackage{diagbox}
\usepackage{cancel}
\usepackage{makecell}
\usepackage[normalem]{ulem}
\usepackage{slashed}
\usepackage{framed}
\usepackage{needspace}

\newcommand\hatslashed[1]{{\hat{#1}\mathllap{\slashed{#1}}}}
\newcommand\tildeslashed[1]{{\tilde{#1}\mathllap{\slashed{#1}}}}

\usepackage[square,numbers,compress]{natbib}
\usepackage{bibentry}

\usepackage[hidelinks, linkcolor=blue, citecolor=blue, urlcolor=blue]{hyperref}

\usepackage{tikz}
\usetikzlibrary{arrows}
\usetikzlibrary{calc}
\usetikzlibrary{decorations.pathreplacing,calligraphy}

\renewcommand{\theequation}{\arabic{section}.\arabic{equation}}

\newcommand{\F}{{\cal F}}

\newcommand{\V}{{\cal V}}

\newcommand{\dd}{\text{d}}

\renewcommand{\Re}{{\rm Re}\,}

\newcommand{\vol}{{\mathcal V}}
\newcommand{\ep}{e^\phi}

\newcommand{\one}{{1\hspace{-3.2pt}\mathrm I}}

\newcommand{\tr}{\mathrm{Tr}}
\newcommand{\ap}{\alpha^\prime}

\newcommand{\vev}[1]{\langle#1\rangle}

\def\im{\mbox{Im}\, }

\newcommand{\overbar}[1]{\mkern 1.5mu\overline{\mkern-1.5mu#1\mkern-1.5mu}\mkern 1.5mu}

\renewcommand{\and}{\mbox{and}}

\usepackage{xcolor}

\newcommand{\Mp}{M_{\rm p}}
\newcommand{\Ms}{M_{\rm s}}
\newcommand{\gs}{g_{\rm s}}

\newcommand{\half}{\frac{1}{2}} 

\renewcommand{\bm}{\boldmath} 

\def\be{\begin{equation}}
\def\ee{\end{equation}}
\def\bea{\begin{eqnarray}}
\def\eea{\end{eqnarray}}
\def\bes{\begin{subequations}}
\def\ees{\end{subequations}}

\definecolor{TC}{HTML}{629202}
\definecolor{AH}{HTML}{d91f05}
\definecolor{JS}{HTML}{5F9EA0}
\definecolor{BF}{HTML}{f903d7}

\def\jac(#1,#2){
\begin{bsmallmatrix}
#1\cr 
#2\cr
\end{bsmallmatrix}}

\newcommand\supsetsim{\mathrel{\ooalign{\raise0.2ex\hbox{$\supset$}\cr\hidewidth\raise-0.9ex\hbox{\scalebox{0.9}{$\sim$}}\hidewidth\cr}}}

\counterwithin*{equation}{section}

\renewcommand{\baselinestretch}{1.2}

\begin{document}

\thispagestyle{empty}

\rightline{}

\begin{center}
{\huge {\bf The String Theory Photoverse}\\[12pt]}
\bigskip
\bigskip
{\bf Thibaut Coudarchet}\footnote{coudarchet@thphys.uni-heidelberg.de},
{\bf Arthur Hebecker}\footnote{a.hebecker@thphys.uni-heidelberg.de},
{\bf Joerg Jaeckel}\footnote{jjaeckel@thphys.uni-heidelberg.de}\\
{\bf and Jonathan Steiner}\footnote{steiner@thphys.uni-heidelberg.de}

\bigskip
\vspace{0.2cm}
{\it Institute for Theoretical Physics, Heidelberg University,\\ Philosophenweg 16\,}\&\,{\it 19, 69120 Heidelberg, Germany }\\[5pt]

\bigskip
\bigskip
\end{center}

\begin{center}
\begin{minipage}[h]{15.0cm} 
String theory compactifications come with numerous $U(1)$ factors, implying the presence of many hidden photons in the low-energy EFT. One may call this the ``string photoverse''. We argue that, generically, these hidden photons are massless and do not couple to any light dark current such that, naively, kinetic mixing with the Standard Model is unobservable. The leading interactions of these ``superhidden'' photons are then dimension-6 dipole operators which couple them to quarks or leptons and the Higgs field. This induces magnetic and electric dipole moments with respect to both the superhidden photons as well as, through kinetic mixing, to the Standard Model photon. We derive these couplings by dimensionally reducing the fermionic action of 7-branes realizing the Standard Model: In the first step to 6d theories on intersection curves and then, in the presence of fluxes, to our 4d chiral EFT. We analyze how experiments and observations can employ this effect to place lower bounds on the string scale, which is relevant for compactifications with very large volumes. Finally, we briefly discuss how supersymmetry implies the presence of relatively light photinos and hence an accompanying ``photinoverse'', which may be observed via renormalizable mixing effects.
\end{minipage}
\end{center}

\newpage

\pagestyle{plain}
\renewcommand{\thefootnote}{\arabic{footnote}}
\setcounter{footnote}{0}

\renewcommand{\baselinestretch}{1.5}


\tableofcontents

\section{Introduction}

The low-energy limit of string theory compactifications generically contains relatively light particles, opening up the possibility of experimentally testing the theory and constraining models (see~\cite{Jaeckel:2010ni,Marsh:2015xka,Agrawal:2021dbo,Antypas:2022asj,Adams:2022pbo,Antel:2023hkf} for a selection of reviews). A prime example is provided by axions. They constitute what is known as the string theory \emph{axiverse} \cite{Conlon:2006tq,Svrcek:2006yi,Arvanitaki:2009fg,Acharya:2010zx,Cicoli:2012sz} which has been explored in large ensembles of string compactifications, see e.g. \cite{Cicoli:2012aq,Higaki:2012ar,Gao:2013rra,Honecker:2013mya,Hebecker:2014gka,Choi:2014uaa,Acharya:2015zfk,Demirtas:2018akl,Halverson:2019cmy,Mehta:2020kwu,Broeckel:2021dpz,Cicoli:2021tzt,Cicoli:2021gss,Cicoli:2022fzy,Gendler:2023kjt,Dimastrogiovanni:2023juq,Agrawal:2024ejr,Petrossian-Byrne:2025mto,Loladze:2025uvf,Benabou:2025kgx,Cheng:2025ggf,Leedom:2025mlr,Reig:2025dqb}.

Another generic feature of the low-energy Effective Field Theories (EFT) arising from string theory is the presence of multiple gauge sectors. These can be spontaneously broken or, in the non-abelian case, undergo confinement already at high energies. Yet, a significant fraction are expected to remain light. Our focus in this paper is on the arguably more generic abelian case (for studies of non-abelian hidden sectors see e.g.~\cite{Faraggi:2000pv,Acharya:2017szw,Dienes:2016vei,Halverson:2016nfq,Halverson:2018olu,Halverson:2020xpg}). As we will review momentarily, abelian hidden sectors are truly ubiquitous, such that the term string theory \emph{photoverse} seems appropriate.

These hidden $U(1)$s arise mainly in two ways: On the one hand, from spacetime-filling branes located away from the Standard Model (SM) sector. On the other hand, from the reduction of RR $p$-form fields on $(p\!-\!1)$-cycles. Our focus will be on type IIB orientifold compactifications, in which case the key players are D3-brane photons, originating in single D3-branes,\footnote{
In the context of the classic type IIB moduli stabilization scenarios \cite{Kachru:2003aw, Balasubramanian:2005zx} with anti-D3 uplift, single D3s would be attracted by and annihiliate with the anti-brane. However, given the serious control problems of the anti-D3 uplift \cite{Carta:2019rhx, Gao:2020xqh, Junghans:2022exo, Gao:2022fdi, Junghans:2022kxg, Hebecker:2022zme, Schreyer:2022len, Schreyer:2024pml, Moritz:2025bsi}, alternative uplifts may appear more natural, making single D3s a generic feature.
} and RR photons, coming from the $C_{(4)}$-potential dimensionally reduced on 3-cycles. It is well-known that a priori massless $U(1)$s can be lifted through a Stückelberg mechanism mediated by 2-form fields. However, this happens neither for our specific case of 3-cycle RR photons nor for the D3-brane photons: In both cases the potentially dangerous 4d terms are schematically of the form $F\wedge B_{(2)}$ or $F\wedge C_{(2)}$. However, the orientifold action projects out the 4d zero modes of these 2-form potentials.\footnote{Note that this is different from the situation with D7-branes or D3-branes at a singularity, where various mechanisms making the U(1)s massive exist.
}

We conclude that a string theory photoverse, containing many hidden photons, is a generic prediction. The subject of hidden photons from string theory has, of course, been extensively studied, cf.~e.g.~\cite{Goodsell:2009pi,Goodsell:2009xc,Arvanitaki:2009hb,
Bullimore:2010aj, Cicoli:2011yh, Camara:2011jg, Marchesano:2014bia, Bauer:2018onh, Anastasopoulos:2020xgu, Hebecker:2023qwl,
Sheridan:2024vtt} (see \cite{Fabbrichesi:2020wbt} for a non-stringy, phenomenological review). The focus has largely been on effects arising from a small mass or light charged fields, which become visible through kinetic mixing.\footnote{A brief discussion of cosmological effects relevant to the massless case appears in \cite{Sheridan:2024vtt}. There has also been significant work on indirect effects related to photini (see e.g.~\cite{Ibarra:2008kn, Arvanitaki:2009hb, Goodsell:2011wn}).
} However, the hidden photons in string theory are generically massless and not coupled to any light dark current. Hence, they do not interact with the SM at the renormalizable level. Such \emph{superhidden} photons are the main constituents of the string theory photoverse and, in what follows, they are the central subject of our interest.

The most direct interactions of superhidden photons with the SM come from the dimension-six operator \cite{Dobrescu:2004wz} 
\begin{equation}
    {\mathcal{L}}_{\rm dipole}=-\frac{1}{2}\frac{v_{\rm h}}{\Lambda^2}\,X_{\mu\nu}\,\overline\psi^{\,i}\sigma^{\mu\nu}\left(d^{\rm M}_{ij}+id^{\rm E}_{ij}\gamma^{5}\right)\psi^j\,,
    \label{eq:dipole_intro}
\end{equation}
where $X_{\mu\nu}$ is the superhidden photon field strength. To respect the $SU(2)_{\rm L}\times U(1)_Y$ gauge invariance a Higgs insertion is needed, hence the presence of the Higgs vev $v_{\rm h}$ and a suppression by $\Lambda^2$,  with $\Lambda$ some energy scale. The hermitian coefficient-matrices $d_{ij}^{\rm M}$ and $d_{ij}^{\rm E}$ parametrize the magnetic and electric couplings respectively between flavors $i$ and $j$. Such an operator has a rich phenomenology \cite{Dobrescu:2004wz,Fabbrichesi:2020wbt} and is constrained by a variety of observations and experiments as we will review and discuss in sect.~\ref{sec:bounds}. Stringent constraints arise from astrophysical observations, notably energy loss in stars and supernovae~\cite{Dobrescu:2004wz,Fabbrichesi:2020wbt,Camalich:2020wac,Carenza:2019pxu}. In presence of flavor or CP violation, muon decays~\cite{Dobrescu:2004wz,Fabbrichesi:2020wbt} as well as the electric dipole moment of the electron also have very high sensitivity. To a lesser extent but with good prospects from experimental improvements, the operator is also constrained from the spin-dependent potentials that it induces~\cite{Dobrescu:2006au,Cong:2024qly,Heckel:2013ina,Terrano:2015sna,Almasi:2018cob}.

A main purpose of this paper is to analyze precisely how this 4d dipole operator arises from type IIB Calabi--Yau orientifolds, both for brane photons and RR photons \cite{Arvanitaki:2009hb,Camara:2011jg}. This will allow us to probe and constrain string models from the experimental bounds mentioned above. Our approach is 10d supergravity based. We will compare with a related worldsheet analysis in \cite{Anastasopoulos:2020xgu} below. The setup we have in mind is that of a large volume compactification with the SM realized on intersecting stacks of D7-branes. Our starting point is then the fermionic action for D-branes \cite{Cederwall:1996pv,Cederwall:1996ri,Bergshoeff:1996tu,deWit:1998tk,Grana:2002tu, Marolf:2003vf, Marolf:2003ye, Martucci:2005rb}, which contains fermion bilinears coupled in particular to $H_{(3)}$ and $F_{(5)}$. This is the origin of our 4d dipole operator. Indeed, the brane photons induce a tail for the $B_{(2)}$ and $C_{(2)}$ fields. The former implies non-zero $H_{(3)}$ near the SM stack and thus couples to the fermions. For the RR photons the coupling is more straightforward since they directly induce $F_{(5)}$. Our strategy is to first reduce the 8d worldvolume action to get a 6d EFT for the fermion modes that live on the 7-brane intersection curves. Then we reduce to 4d by considering magnetized branes to induce a chiral spectrum and generate our dipole coupling. A Pati--Salam toy model is used for illustration. We find a suppression of the dimension 6 operator by a scale $\Lambda=\alpha\Ms$, with $\Ms$ the string scale and $\alpha$ a numerical factor to be defined later. We argue that the effect is IR dominated and that $\alpha$ may be somewhat less than unity.

Combining the experimental bounds with our result for the dipole operator, we derive lower limits for the string scale (cf.~sect.~\ref{sec:pheno}). These bounds can be significantly stronger than those from LHC and 5th force experiments, making them highly relevant and allowing in principle for near future discovery. However, such an optimistic conclusion is justified only if we are agnostic about the specific mechanism by which the volume modulus is stabilized. If, by contrast, we consider more explicit setups like the Large Volume Scenario (LVS) \cite{Balasubramanian:2005zx,Conlon:2005ki}, consistency with a SUSY breaking scale of at least $\sim 1$ TeV puts constraints on the string scale that are sharper or comparable to those presently derivable using the dipole operator. The string scale can be lowered if one supplements an LVS-type moduli stabilization scheme with a SUSY breaking sector coupled strongly to the SM. This possibility is, however, limited by the \emph{F-term problem} \cite{Hebecker:2019csg} as we explain in sect.~\ref{sec:pheno}. In spite of this limitation, models of this type are significantly constrained by the dipole operator already at present experimental sensitivity.

Beyond the dipole operator, supersymmetry and its breaking can make massless hidden photons visible through the accompanying photini~\cite{Ibarra:2008kn, Arvanitaki:2009hb, Goodsell:2011wn} and their mixing effects with neutralinos. We briefly venture into this \emph{photinoverse} in sect.~\ref{sec:photinoverse}, but leave a more detailed study of this exciting field to future work.

The structure of the paper is as follows: In sect.~\ref{sec:VisSecCoup} we describe the geometrical setup that we consider in our study and the origin and couplings of superhidden photons in string theory. In sect.~\ref{sec:profiles} we evaluate the profiles of $H_{(3)}$ and $F_{(3)}$ induced by photons on a single D3-brane, as well as the $F_{(5)}$ bulk profile relevant for RR photons. Sect.~\ref{sec:6d} is devoted to the dimensional reduction of the SM D7-branes worldvolume theory to 6d on intersection curves. Sect.~\ref{sec:4d} introduces a Pati--Salam toy model and treats the reduction to 4d in situations where the branes are magnetized. This allows us to explain how the crucial dipole operator for chiral SM fields arises. In sect.~\ref{sec:bounds} we review different experimental constraints on the dipole operator and in sect.~\ref{sec:pheno} we discuss what they imply for string theory constructions. In sect.~\ref{sec:photinoverse} we consider effects of the photini, i.e.~the superpartners of the hidden photons. Conclusions and outlook are discussed in sect.~\ref{sec:conclusion}. Three appendices supplement the paper: Appendix~\ref{app:conv} summarizes conventions for the Hodge star, gamma matrices and spinors. Appendix~\ref{app:Red} provides details concerning the dimensional reduction to 6d and 4d, treating the special case of a magnetized torus explicitly. Finally, appendix~\ref{app:Fterm} presents our estimation of the $F$-terms, required in sect.~\ref{sec:photinoverse}.

\section{Origin and interactions of superhidden photons}
\label{sec:VisSecCoup}

In this section we describe the setup relevant for the rest of the paper: A type IIB orientifold compactification with the SM realized on a stack of branes. We will review two types of superhidden photons that generically appear in this setup: D3-brane photons and closed-string (RR) photons. We then explain how 4d dipole couplings allow the superhidden photons to interact with SM fermions. 

\subsection{Setup and geometric picture}
\label{sec:setup}

\begin{figure}
    \centering
    \includegraphics[scale=0.85]{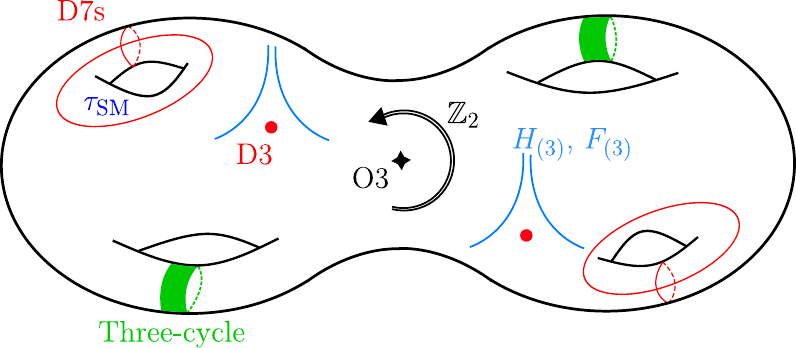}
    \caption{Illustration of our geometric setup. We consider type IIB orientifolds with D3/D7-branes. The SM is realized on intersecting D7-branes on a 4-cycle with volume $\tau_{\rm SM}$. An isolated spacetime-filling D3-brane sits somewhere in the internal space. The D3-brane worldvolume field strength sources $H_{(3)},\,F_{(3)}$ profiles which in turn couple to matter living on the SM D7s. Additional 4d $U(1)$ gauge bosons arise from the dimensional reduction of $C_{(4)}$ on three-cycles (depicted in green).}
    \label{fig:geom}
\end{figure}

We consider type IIB string theory compactifications on Calabi--Yau (CY) orientifolds with fluxes and O3/O7-planes \cite{Giddings:2001yu, Grana:2005jc, Denef:2008wq}.
Apart from assuming a large overall volume $\V$ to ensure EFT control, our discussion is generic. The study of more specific constructions is postponed to section~\ref{sec:LVS}. We take the SM to be realized on intersecting stacks of D7-branes wrapping (relatively small) four-cycles,  collectively denoted $\tau_{\rm SM}$. This setting allows for spacetime filling D3-branes, pointlike in the internal space. Together with the three-form flux, they contribute to cancelling the O-plane induced D3-tadpole. Hence, one expects them to be generically present. While their location in the CY is a flat direction at the level of the GKP analysis~\cite{Giddings:2001yu}, subleading effects will fix these moduli. We focus on branes sitting at smooth points of the CY. Figure \ref{fig:geom} summarizes the situation.

In this framework we expect two types of superhidden photons:
\begin{itemize}
    \item The $U(1)$ worldvolume gauge theory of each D3-brane contains 
    a massless photon without light matter charged under it. The only fields potentially relevant for giving these photons a St\"uckelberg mass are the 4d two-forms coming from the 10d fields
    $B_{(2)}$ and $C_{(2)}$.
    However, since these 4d two-forms are removed by the orientifolding, all U(1)s remain massless.
    \item The KK reduction 
    of $C_{(4)}$ generates 4d photons $ V_ {(1)}^\kappa$, $ U_{(1)\kappa}$ associated with the basis three-forms $\alpha_\kappa$ and $\beta^\kappa$ ($\kappa=1,\dots,h^{(2,1)}_+$) even under the orientifold involution \cite{Grimm:2004uq}:
    \begin{equation}
C_{(4)}=V_{(1)}^\kappa(x)\wedge\alpha_\kappa+U_{(1)\kappa}(x)\wedge\beta^\kappa+\cdots
    \end{equation}
    Note that half of these degrees of freedom are removed by the self-duality constraint on $F_{(5)}$. For obvious reasons, the photons arising in this way are known as ``RR photons''~\cite{Arvanitaki:2009hb,Camara:2011jg}. Since there are no light states charged under them, they are also superhidden.\footnote{Additional massive RR photons related to torsional cohomology classes have been studied in~\cite{Camara:2011jg}.}
\end{itemize}

\subsection{Superseeking the superhidden photons}
\label{sec:superhidden}

Massless hidden photons (see \cite{Fabbrichesi:2020wbt} for a review) are usually probed through the hypercharge portal, i.e.~through a 4d kinetic mixing between the hypercharge U(1) and the hidden photon~\cite{Okun:1982xi,Holdom:1985ag,Foot:1991kb}. The 4d Lagrangian contains renormalizable terms like
\begin{equation}
    \mathcal{L}_{4\rm d}\supset-\frac{1}{4}F_{\mu\nu}F^{\mu\nu}-\frac{1}{4}X_{\mu\nu}X^{\mu\nu}-\frac{\epsilon}{2}F_{\mu\nu}X^{\mu\nu}+e_{\rm SM}J^\mu_{\rm SM}A_\mu+e_{\rm DS}J^\mu_{\rm DS}X_\mu\,.
\end{equation}
Here $F_{\mu\nu}$ is the field strength of $A_\mu$ which couples to the SM matter current $J^\mu_{\rm SM}$ with coupling $e_{\rm SM}$; $X_{\mu\nu}$ is the field strength of $X_\mu$ which couples to the dark current $J^\mu_{\rm DS}$ with coupling $e_{\rm DS}$; and $\epsilon$ sets the magnitude of the kinetic mixing between the two field strengths. The mixing term can be eliminated \cite{Holdom:1985ag} by the field redefinition\footnote{Note that a different field redefinition can be performed \cite{Fabbrichesi:2020wbt}, resulting in the hidden photon to also be coupled to the SM matter current, but because this field redefinition introduces a further coupling, it is more convenient to work with \eqref{eq:rotation}.}
\begin{equation}
\label{eq:rotation}
    \begin{pmatrix}
        A_\mu'\\
        X_\mu'
    \end{pmatrix}=
    \begin{pmatrix}
        \sqrt{1-\epsilon^2} & 0\\
        \epsilon & 1
    \end{pmatrix}
    \begin{pmatrix}
        A_\mu\\
        X_\mu
    \end{pmatrix}.
\end{equation}
One can then identify $A_\mu'$ as our actual hypercharge photon, which couples to both $J^\mu_{\rm SM}$ and $J^\mu_{\rm DS}$, while $X_\mu'$ plays the role of the hidden photon and couples to the dark current only. When part of the dark sector is light, the coupling proportional to $\epsilon$ between the dark current and $A_\mu'$ is known as \emph{millicharge} coupling.  

On the other hand, when no light state at all couples to the massless hidden photon, $\epsilon$ is unobservable from the 4d renormalizable Lagrangian, thus rendering the photon \emph{superhidden}. 

A direct coupling to the SM can then only occur through higher-dimension operators. A dimensions-six Lagrangian, involving the Higgs for chirality reasons, is the strongest effect one can hope for \cite{Dobrescu:2004wz},
\begin{equation}
\label{eq:dipoleint}
    {\mathcal{L}}_{\rm dipole}=-\frac{1}{2}\frac{v_{\rm h}}{\Lambda^2}X_{\mu\nu}\overline\psi^{\,i}\sigma^{\mu\nu}\left(d^{\rm M}_{ij}+id^{\rm E}_{ij}\gamma^{5}\right)\psi^j\,.
\end{equation}
Here $v_{\rm h}$ is the Higgs vev, $\sigma^{\mu\nu}\equiv\frac{i}{2}[\gamma^\mu_{\mathsf P},\gamma^\nu_{\mathsf P}]$, the matrices $d^{\rm M,E}_{ij}$ are both hermitian and parameterize the strength of the magnetic and electric dipole moments, and $\Lambda$ is the energy scale suppressing the operator. Note that, for easier comparison with the phenomenological literature, the 4d gamma matrices above differ form those used in the rest of the paper. They are related by $\gamma^\mu=i\gamma^{\mu}_{\mathsf P}$. The dipole operator \eqref{eq:dipoleint} is the main focus of this paper as it represents the strongest direct experimental probe of superhidden photons.

Note that the field rotation \eqref{eq:rotation} gives
\begin{equation}
    {\mathcal{L}}_{\rm dipole}\simeq-\frac{1}{2}\frac{v_{\rm h}}{\Lambda^2}\left(X'_{\mu\nu}-\epsilon F'_{\mu\nu}\right)\overline\psi^{\,i}\sigma^{\mu\nu}\left(d^{\rm M}_{ij}+id^{\rm E}_{ij}\gamma^{5}\right)\psi^j\,.
    \label{eq:dipole}
\end{equation}
The terms involving $X'_{\mu\nu}$ induce dark electric and magnetic dipole moments. By contrast, the terms involving the hypercharge field strength $F_{\mu\nu}'$ induce standard dipole moments, thereby providing additional, indirect means to observe the mixing parameter $\epsilon$. Phenomenological consequences of this dipole Lagrangian together with experimental bounds will be reviewed and discussed in sects.~\ref{sec:bounds} and \ref{sec:pheno}. Before turning to this subject, we now describe in detail how such 4d dipole interactions arise from the 10d theory.

\section{Photon localization and profiles}
\label{sec:profiles}

\subsection{Brane photons}

In our setup, depicted in fig.~\ref{fig:geom}, the photon living on the D3-brane sources $B_{(2)}$ and $C_{(2)}$ profiles, which in turn induce non-trivial $H_{(3)}$ and $F_{(3)}$ field strengths at the location of SM branes. These field strengths affect the fermionic part of the SM brane action through terms of the type $\overline\Theta \slashed{H}_{\!(3)}(\dots)\Theta$ and $\overline\Theta\slashed{F}_{\!(3)}(\dots)\Theta$, where $\Theta$ is a 7-brane spinor to be defined later. As a result, dipole interactions of the form \eqref{eq:dipoleint} are induced if two indices of the 3-form field strength point along the non-compact 4d spacetime directions. As a first step we thus evaluate the profiles of $H_{(3)}$ and $F_{(3)}$ induced by the D3-brane photon.

Our starting point is a total bosonic action $S$ split into the bulk type IIB supergravity action and the action of the D3-brane,
\begin{equation}
    S=S_{\rm IIB}+S_{\rm D3}\,.
\end{equation}
In the conventions of \cite{Myers:1999ps,Martucci:2005rb} the bulk action reads,
\begin{align}
\begin{split}
    S_{\rm IIB}=&\frac1{2\kappa^2}\int e^{-2\phi}\bigg(R\star1+4\dd\phi\wedge\star \dd\phi-\frac12 H_{(3)}\wedge\star H_{(3)}\bigg)\\
&+\frac1{4\kappa^2}\int \bigg(-F_{(1)}\wedge\star F_{(1)}-F_{(3)}\wedge\star F_{(3)}-\frac1{2}F_{(5)}\wedge\star F_{(5)}\bigg)\label{eq:IIBaction}\\
&+\frac1{4\kappa^2}\int  \dd C_{(2)}\wedge H_{(3)}\wedge \bigg(C_{(4)}+\frac12B_{(2)}\wedge C_{(2)}\bigg)\,,
\end{split}
\end{align}
with
\begin{equation}
F_{(1)}=\dd C_{(0)}\,,\,
F_{(3)}=\dd C_{(2)}+C_{(0)}H_{(3)}\,,\,
F_{(5)}=\dd C_{(4)}+C_{(2)}\wedge H_{(3)}\text{ and }F_{(5)}=-\star F_{(5)}\,.
\end{equation}
The D3-brane action is further split into the DBI action and the Chern--Simons part, $S_{\rm D3}=S_{\rm DBI}+S_{\rm CS}$, with
\begin{equation}
\label{eq:DBI}
    S_{\rm DBI}=-\mu_{\rm D3}\int_{\rm D3}e^{-\phi}\sqrt{-\det(g+\F_{(2)})}\quad \mbox{and}\quad S_{\rm CS}=\mu_{\rm D3}\int_{\rm D3}
    e^{\,\F_{(2)}}\,
    \sum_q C_{(q)} +\cdots
\end{equation}
Here, $\F_{(2)}\equiv B_{(2)}+2\pi\alpha'F_{(2)}$ and $F_{(2)}=\dd A_{(1)}$ is the D3-brane worldvolume field strength. The dots in $S_{\rm CS}$ contain curvature corrections and a pullback to the worldvolume is understood whenever appropriate. We also use $2\kappa^2=(2\pi)^7\alpha^{\prime 4}$, $\mu_{\rm D3}^{-1}=(2\pi)^3\alpha^{\prime 2}$ \,\cite{Giddings:2001yu}.

We want to evaluate the profiles for $F_{(3)}$ and $H_{(3)}$ sourced by a hidden-photon field strength $\F_{(2)}$ on the D3-brane. For this, we solve the 10d equations of motion at large volume (implying weak warping) and for constant axio-dilaton. We disregard the $|F_{(5)}|^2$ and Chern--Simons terms since, while these terms contain $B_{(2)}$ and $C_{(2)}$, their effects are higher order at large volume. The variaton of the action then reads
\begin{align}
\begin{split}
    \delta S=&\int\delta B_{(2)}\wedge\bigg(\frac{e^{-2\phi}}{2\kappa^2}\dd\star H_{(3)}+\frac{C_{(0)}}{2\kappa^2}\dd\star F_{(3)}-\mu_{\rm D3}e^{-\phi}(\star_4\mathcal F_{(2)})\wedge\delta_6(y)\\
    &\qquad\qquad\qquad\qquad\qquad\qquad+\mu_{3}C_{(2)}\wedge\delta_6(y)+\mu_{\rm D3}C_0\mathcal F_{(2)}\wedge\delta_6(y)\bigg)\\
    +&\int\delta C_{(2)}\wedge\bigg(\frac1{2\kappa^2}\dd\star F_{(3)}+\mu_{\rm D3}\mathcal F_{(2)}\wedge\delta_6(y)\bigg)\\
    +&\int\delta A_{(1)}\wedge\bigg(-\dd_4\star_4\mathcal F_{(2)}+C_0\dd_4\mathcal F_{(2)}+\dd_4C_{(2)}\bigg)\wedge\delta_6(y)\mu_{\rm D3}2\pi\ap e^{-\phi}\,.
\end{split}
\end{align}
Here, we use $y$ to denote the internal coordinate and let the D3-brane be localized at $y=0$, characterized by the 6-form $\delta_6(y)$. Choosing the gauge $B_{(2)\mu\nu}=C_{(2)\mu\nu}=0$ and using the scalar Calabi--Yau Green's function $G$,
\begin{equation}
(\dd_6\dd_6^\dagger+\dd_6^\dagger\dd_6)G(y_1,y_2)=\delta_0(y_1-y_2)=\star_6\,\delta_6(y_1-y_2)\,,
\end{equation}
we find
\begin{equation}
H_{(3)}=-4\pi\ap e^{\phi}\kappa^2\mu_{\rm D3} F_{(2)}\wedge\dd_6 G\,,\qquad F_{(3)}=-4\pi\ap\kappa^2\mu_{\rm D3} \star_4 F_{(2)}\wedge\dd_6 G\,.
\end{equation}
Equivalently, we have
\begin{equation}
B_{(2)}=-4\pi\ap e^{ \phi}\kappa^2\mu_{\rm D3} A_{(1)}\wedge\dd_6 G\,,\qquad C_{(2)}+C_{(0)}B_{(2)}=-4\pi\ap\kappa^2\mu_{\rm D3} \tilde{A}_{(1)}\wedge\dd_6 G\,,
\end{equation}
where $\tilde{A}_{(1)}$ is the gauge potential for the magnetic field strength $\star_4 F_{(2)}$.

For later use we make the profile near the brane, where the CY can be approximated by flat space (i.e. $g_{mn}\sim\eta_{mn}$), explicit. We then have $G=+1/(4\pi^3 r^4)$ with $r^2\equiv g_{mn}y^m y^n$. This gives
\begin{equation}
\label{eq:F3H3}
H_{(3)}=32\pi^2\sqrt{\alpha'}\left(\frac{\sqrt{\alpha'}}{r}\right)^5g_{\rm s}F_{(2)}\wedge\dd_6 r\,,\quad F_{(3)}=32\pi^2\sqrt{\alpha'}\left(\frac{\sqrt{\alpha'}}{r}\right)^5\star_4 \!F_{(2)}\wedge\dd_6 r\,.
\end{equation}

\subsection{RR photons}

We now proceed by repeating the analysis above for RR photons, i.e.~by determining the corresponding $F_{(5)}$ bulk profile. The required dimensional reduction of type IIB orientifold geometries with D3/D7-branes has been worked out in \cite{Grimm:2004uq}. The analysis is essentially the same as for an $\mathcal N=2$ reduction on a Calabi--Yau, but dropping fields which are odd under orientifolding. We briefly recall the relevant details in what follows.

One first defines a symplectic basis of $H^{(3)}_+$, i.e.~harmonic three-forms $\alpha_\kappa,\,\beta^\kappa$ with 
\be
\int_{\rm CY}\alpha_\kappa\wedge\beta^\lambda=\delta_\kappa^\lambda\,,\qquad\int_{\rm CY}\alpha_\kappa\wedge\alpha_\lambda=\int_{\rm CY}\beta^\kappa\wedge\beta^\lambda=0\,,\label{eq:symplectic}
\ee
where $\kappa,\,\lambda=1,\dots,\mathrm{dim}\big(H^{(3)}_+\big)$. As the Hodge star preserves the sign under the orientifold involution, it is possible to find real matrices ${A_\kappa}^\lambda,\,B_{\kappa\lambda},\,C^{\kappa\lambda}$ and ${D^\kappa}_\lambda$ such that
\be
\star_6\,\alpha_\kappa={A_\kappa}^\lambda\alpha_\lambda+B_{\kappa\lambda}\beta^\lambda\quad\text{ and }\quad \star_6\beta^\kappa =C^{\kappa\lambda}\alpha_\lambda+{D^\kappa}_\lambda\beta^\lambda\,.\label{eq:abstar}
\ee
As already mentioned in sect.~\ref{sec:setup}, the KK ansatz for $F_{(5)}$ reads 
\be
F_{(5)}=\mathsf F_{(2)}^\kappa\wedge\alpha_\kappa+\mathsf G_{(2)\kappa}\wedge\beta^\kappa +\cdots\label{eq:F5ans}
\ee
with $\mathsf F_{(2)}^\kappa=\dd V_{(1)}^\kappa$ and $\mathsf G_{(2)\kappa}=\dd U_{(1)\kappa}$ being 4d two-from field strengths. However, due to the self-duality of $F_{(5)}$, $\mathsf F_{(2)}^\kappa$ and $\mathsf G_{(2)\kappa}$ are not independent, but related by a magnetic--electric duality. Given the $A,B,C$ and $D$ matrices, the reduction is straightforward: \eqref{eq:F5ans} is inserted into \eqref{eq:IIBaction} and evaluated using \eqref{eq:symplectic} and \eqref{eq:abstar}. The self-duality is then imposed a posteriori by adding a Lagrange multiplier term $\sim\mathsf F^\kappa_{(2)} \wedge\mathsf G_{(2)\kappa}$ \cite{DallAgata:2001brr}. One finds 
\begin{equation}
S^{\rm RR}_{\rm 4d}=\frac1{4\kappa^2}\int\bigg(C^{-1}_{\kappa\lambda}\mathsf F_{(2)}^\kappa\wedge\star_4\mathsf F_{(2)}^\lambda- (AC^{-1})_{\kappa\lambda}\mathsf F_{(2)}^\kappa\wedge \mathsf F_{(2)}^\lambda\bigg)\,.\label{eq:RR4d}
\end{equation}

Even though this procedure is necessary for the reduction of the action, for the purpose of determining the bulk profile of $F_{(5)}$ one can directly impose the self-duality on the field. This fixes $\mathsf G_{(2)\kappa}$ in terms of $\mathsf F^\kappa_{(2)}$, 
\be
\mathsf G_{(2)\kappa}=-\mathsf F_{(2)}^\lambda\big(AC^{-1})_{\lambda\kappa}+\star_4\mathsf F_{(2)}^\lambda\big(C^{-1})_{\lambda\kappa}\,,
\ee
which translates to
\be
F_{(5)}=\mathsf F_{(2)}^\lambda\wedge\alpha_\lambda+\Big(-\mathsf F_{(2)}^\lambda\big(AC^{-1})_{\lambda\kappa}+\star_4\mathsf F_{(2)}^\lambda\big(C^{-1})_{\lambda\kappa}\Big)\wedge\beta^\kappa\,.
\label{eq:F5self}
\ee
Thus, for any given RR photon, at least two three-cycles contribute: One via the electric and one via the magnetic field strength.

\needspace{5\baselineskip}
\section[Deriving the 6d fermion bilinear coupling from the 10d theory]{\texorpdfstring{Deriving the 6d fermion bilinear coupling from\\[.2cm]
the 10d theory}}
\label{sec:6d}

As noted above, we are interested in a matter sector realized on D7-branes. The 4d dipole couplings in question can involve a D3-brane photon or an RR photon associated with $C_{(4)}$. In the first case, the relevant couplings are of the type $\overline\Theta\slashed H_{\!(3)}(\dots)\Theta$ and $\overline\Theta\slashed F_{\!(3)}(\dots)\Theta$, with $\Theta$ standing for D7-brane-localized fermions. In the second case, the relevant coupling is of the type $\overline\Theta\slashed F_{\!(5)}(\dots)\Theta$.

In the present section, we first rewrite the D7-brane action, which is usually given in 10d spinor notation, in an explicit 8d form. Then we turn to the effectively 6d fermion modes living on 7-brane intersection curves. For clarity, table~\ref{tab:indices} summarizes our index conventions.\footnote{Note also that throughout the paper we assume the external 4d coordinates to be dimensionful with a dimensionless Minkowski metric. By contrast, the internal coordinates are taken to be dimensionless, with a dimensionful metric.}

\renewcommand{\arraystretch}{1.2}
\begin{table}
    \centering
    \begin{tabular}{c|c|c}
       Coordinates  & Indices & Gamma matrices\\\hline\hline
       $x^0,\dots,x^9$  & $M,N,\dots$ & $\hat\Gamma$\\\hline
       $x^0,\dots,x^7$ & $\alpha,\beta,\dots$ & $\Gamma$\\\hline
       $x^0,\dots,x^5$ & $a,b,\dots$ & $\tilde\gamma$\\\hline
       $x^0,\dots,x^3$ & $\mu,\nu,\dots$ & $\gamma$\\\hline
       $x^8,x^9$ and $x^8+ix^9$ & $M',\dots$ and $u$ &\\\hline
       $x^6,x^7$ and $x^6+ix^7$ & $\alpha',\dots$ and $v$ &\\\hline
       $x^4,x^5$ and $x^4+ix^5$ & $a',\dots$ and $z$ &\\
    \end{tabular}
    \caption{Summary of our index conventions in various dimensions. The different symbols used for gamma matrices in 10d, 8d, 6d and 4d are also displayed.}
    \label{tab:indices}
\end{table}

\subsection{Fermionic D7-brane action}

The fermionic action for a single D-brane has been developed in superspace formalism in~\cite{Cederwall:1996pv,Cederwall:1996ri,Bergshoeff:1996tu,deWit:1998tk} and was then worked out in terms of component fields by e.g.~\cite{Grana:2002tu, Marolf:2003vf, Marolf:2003ye, Martucci:2005rb}. We use the conventions of \cite{Martucci:2005rb}, consistently with the bosonic bulk and brane actions given earlier (see appendix \ref{app:conv} for more details). At quadratic order, the fermionic action reads,
\begin{equation}
    S^{\rm (F)}_{\rm D7}=\frac{\mu_{\rm D7}}2\int_{\mathcal D} \dd^{8}\xi e^{-\phi}\sqrt{-\det (g+\mathcal F)}\overline\Theta(1-\Gamma_{\rm D7})[(\tilde M^{-1})^{\alpha\beta}\hat\Gamma_\beta D_\alpha-\Delta]\Theta\,.\label{eq:SD7F}
\end{equation}
Here $\Theta$ stands for a 2-vector containing a pair of 10d Majorana--Weyl fermions $\hat\theta_{1,2}$, each of positive chirality,
\be
\Theta\equiv\begin{pmatrix}\hat\theta_1\\\hat\theta_2\end{pmatrix},\qquad \hat\Gamma_{(10)}\hat\theta_i=+\hat\theta_i\,.
\label{Tdef}
\ee
The index, $\alpha=0,\dots,7$, runs over worldvolume directions while the indices $M,N=0,\dots,9$ characterize the full 10d spacetime. Throughout this paper we work in static gauge, i.e. the brane is located at fixed $x^{8,9}$ and the worldvolume coordinates $\xi^\alpha$ are set equal to the corresponding ten-dimensional $x^\alpha$ ones. Furthermore, as in the previous section, $\mathcal F_{(2)}$ is the standard gauge invariant field strength on the brane: $\mathcal F_{(2)}\equiv B_{(2)}+2\pi\ap F_{(2)}$. The matrix $\tilde M_{\alpha\beta}$ is given by $\tilde M_{\alpha\beta}=g_{\alpha\beta}+\mathcal O(\mathcal F_{(2)})$ and, assuming an adequate vielbein, $\Gamma_{\rm D7} = i\sigma_2\hat\Gamma_{(10)}\hat\Gamma^{\underline9}\hat\Gamma^{\underline8}+\mathcal O(\mathcal F_{(2)})$.\footnote{Comparing \eqref{eq:SD7F} and \eqref{Tdef} it is consistent to set $\Gamma_{\rm D7}=-i\sigma_2\hat\Gamma^{\underline9}\hat\Gamma^{\underline8}+\mathcal O(\mathcal F_{(2)})$.} Here and elsewhere, underlined indices are `flat', which is implemented by an appropriate contraction with a vielbein. Note that $\sigma_2$ and further Pauli matrices appearing below act on the 2-vector of \eqref{Tdef}.  The derivative $D_\alpha$ and the $\Delta$ term depend on 10d fields, 
\begin{align}
    D_\alpha&=\nabla_\alpha+\frac{\ep}{16\cdot5!}\slashed F_{\!(5)}\hat\Gamma_\alpha(i\sigma_2){+\frac1{8}\big(H_{(3)\,\alpha MN}\hat\Gamma^{MN}\sigma_3+\frac{e^\phi}{3!}\slashed F_{\!(3)}\hat\Gamma_\alpha\sigma_1\big)}+\cdots\label{eq:Dalpha}\\
    \Delta&={\frac1{4\cdot3!}\big(\slashed H_{\!(3)}\sigma_3-e^\phi\slashed F_{\!(3)}\sigma_1\big)}+\cdots\label{eq:Delta}
\end{align}
Here the ellipses stand for terms involving only the axio-dilaton. We also use the notation $\slashed A\equiv\hat\Gamma^{M_1\cdots M_n}A_{M_1\cdots  M_n}$. Note that the derivative $D_\alpha$ and $\Delta$ also characterize the infinitesimal SUSY transformations of gravitino and dilatino: $\delta_\varepsilon\psi_M=D_M\varepsilon$ and $\delta_\varepsilon\lambda=\Delta\varepsilon$~\cite{Martucci:2005rb}.

The action \eqref{eq:SD7F} enjoys a local fermionic symmetry known as \emph{kappa--symmetry}. In our context, this is crucial since it reduces the formally $2\times 16=32$ degrees of freedom of $\Theta$ to the $16$ off-shell fermionic degrees of the brane-localized theory. Following the standard SUGRA literature, we gauge-fix the kappa--symmetry by imposing, 
\begin{equation}
    \Theta=\begin{pmatrix}\hat\theta\\
   0\end{pmatrix}.\label{eq:KappaGauge}
\end{equation}
Combining \eqref{eq:SD7F}-\eqref{eq:KappaGauge}, the relevant part of the fermionic D7-brane action takes the form,
\begin{align}
2\mu_{\rm D7}^{-1}\,  S_{\rm D7}=\int_{\mathcal D} \dd^8\xi e^{-\phi}\sqrt{-g}\Bigg[&\overline{\hat\theta}\hat\Gamma^\alpha\nabla_\alpha\hat\theta-\frac 1{24}\Big(\overline{\hat\theta} \slashed H_{\!(3)}{\hat\theta}
    -\ep\overline{\hat\theta}\hat\Gamma^{\underline9}\hat\Gamma^{\underline8}\slashed F_{\!(3)}{\hat\theta}\Big)\nonumber\\
    &+\frac18\Big(\overline{\hat\theta}\hat\Gamma^\alpha\hat\Gamma^{MN}{\hat\theta} H_{(3)\alpha MN}\label{eq:S10eff}
    +\frac{ \ep}6\overline{\hat\theta}\hat\Gamma^{\underline9}\hat\Gamma^{\underline8}\hat\Gamma^\alpha\slashed F_{\!(3)}\hat\Gamma_\alpha{\hat\theta} \Big)\\
   &-\frac{e^\phi}{16\cdot 5!}\overline{\hat\theta}\hat\Gamma^{\underline9}\hat\Gamma^{\underline8}\hat\Gamma^\alpha\slashed F_{\!(5)}\hat\Gamma_\alpha{\hat\theta}\Bigg]+\mathcal O(\mathcal F_{(2)})\nonumber\,.
\end{align}

All of the above applies to a single D7-brane. However, realistic models require brane stacks. Unfortunately, the fermionic counterpart to the Myers action \cite{Myers:1999ps}, i.e.~the non-abelian generalization of \eqref{eq:SD7F}, is still unknown. It is possible in principle to derive the action by T-dualizing the D0-brane action (see \cite{Taylor:1999gq,Taylor:1999pr}) but we shall adopt the approach of \cite{Grana:2002nq,Marchesano:2010bs} and simply promote $\hat \theta$ to an adjoint field. At the same time $\nabla_\alpha$ is promoted to be covariant also with respect to gauge transformations. For the case of a D3-brane, the two approaches have been shown to agree for the lower-order terms \cite{Grana:2002nq}. 

Thus, $\hat\theta$ becomes matrix-valued, all terms in \eqref{eq:S10eff} appear under a trace, and the following terms are added for gauge covariance, 
\begin{equation}
    -\tr\left(i\overline{\hat\theta}\hat\Gamma^\alpha[A_\alpha,\hat\theta]+i\overline{\hat\theta}\hat\Gamma^{M'}[\Phi_{M'},\hat\theta]\right)\,,\quad M'=8,9\,.
\end{equation}
Here $A_\alpha$ is the worldvolume gauge potential and $\Phi^{M'}$ are brane position moduli.

\subsection{Manifestly 8d form}
\label{man8d}

The equations \eqref{eq:SD7F} and \eqref{eq:S10eff} are formulated in a mixture of 8d and 10d language. The goal of this subsection is to arrive at a manifestly 8d action including non-abelian effects.

Since in eight dimensions Majorana spinors exist, it is possible to find a set of gamma matrices $\Gamma^{\underline\alpha},\,\underline\alpha=0,\dots,7$ which are purely imaginary. For an 8d spinor $\theta$, the Majorana condition then reads $\theta=\theta^*$. From $\Gamma^{\underline\alpha}$ it is possible to construct purely real 10d Gamma matrices via the decomposition (see e.g.~\cite{Tomasiello:2022dwe, Ortin:2015hya}),
\begin{equation}
    \hat\Gamma^{\underline\alpha}=\Gamma^{\underline\alpha}\otimes\sigma_2,\quad\hat\Gamma^{\underline8}=\Gamma_{(8)}\otimes\sigma_2\quad\text{and}\quad\hat\Gamma^{\underline9}=-\one_{16}\otimes \sigma_1\,.\label{eq:Gamma10to8}
\end{equation}
In this particular basis the Majorana condition also reads $\hat\theta^*=\hat\theta$. Additionally, $\hat\Gamma_{(10)}$ has the particularly nice form
\begin{equation}\hat\Gamma_{(10)}=\hat\Gamma^{\underline0\cdots\underline9}=\one_{16}\otimes\sigma_3\,.
\end{equation}
This implies that a 10d Majorana--Weyl spinor of positive chirality is written like $\hat\theta=(\theta\quad0)^T$ with $\theta$ being an 8d Majorana spinor. Using this fact and together with \eqref{eq:Gamma10to8}, the Yang--Mills part of the action is easily determined to be,
\begin{equation}
    S^{\rm YM}_{\rm 8d}=\frac{\mu_{\rm D7}}2\int_{\mathcal D} \dd^{8}\xi e^{-\phi}\sqrt{-g}\tr\big\{\overline\theta\Gamma^\mu \nabla_\mu\theta-i\overline\theta P_+[\varphi,\theta]+i\overline\theta P_-[\overline\varphi,\theta]\big\}\,,\label{eq:SYM8G}
\end{equation}
where 

\begin{equation}
P_\pm\equiv(\one\pm\Gamma_{(8)})/2\quad \text{and} \quad \varphi\equiv\Phi^8+i\Phi^9\,. 
\end{equation}
Note that there is a residual $U(1)$ symmetry with actions $\varphi\rightarrow e^{i\alpha}\varphi$ and $\theta\rightarrow\exp(i\alpha\Gamma_{(8)}/2)\theta$ originating from $SO(2)$ rotations in the local $\underline 8\underline9$-plane.

The action in \eqref{eq:SYM8G} is supplemented by the flux-dependent terms from \eqref{eq:S10eff}, which have to be rewritten in explicit 8d language. In doing so, various cancellations arise. Crucially, some of those are due to the Majorana condition $\theta=\mathcal B^{-1}\theta^*$. Using this condition, any spinor bilinear can be written as $\theta^T\Gamma\theta$, with $\Gamma$ a generic combination of gamma matrices. It vanishes if $\Gamma$ is symmetric\footnote{In our case $\Gamma$ is of the form $\mathcal (B^{-1})^\dagger\Gamma^0\Gamma^{\alpha_1\cdots\alpha_n}\Gamma_{(8)}^\eta$ with $(B^{-1})^\dagger\Gamma^0$ coming from the Dirac bar and $\eta=0,1$ to indicate the presence of a 8d chirality matrix or not. Then we have $\Gamma^T=(-1)^{\frac{n(n+2\eta-1)}{2}}\Gamma$ upon using the following properties: $\mathcal B^{-1}=\mathcal B^\dagger$ and $\mathcal B^T=\mathcal B$ \cite{Freedman:2012zz}.}. Combining all surviving terms we find,
\begin{align}
    S^{\rm flux}_{\rm 8d}=\frac{\mu_{\rm D7}}{2}\int_{\mathcal D} \dd^8\xi\sqrt{-g}e^{-\phi}\tr\bigg\{&\frac1{12}
    \overline\theta\Gamma^{\alpha\beta\gamma} H_{(3)\alpha\beta\gamma}\theta+\frac{1}{4}\overline\theta\Gamma^{\alpha\beta}\left(-H_{(3)\alpha\beta \underline u}P_-+H_{(3)\alpha\beta\underline{\bar u}}P_+\right)\theta\nonumber\\
    &-\frac{ie^\phi}{4}\overline\theta\Gamma^{\alpha\beta}\left(F_{(3)\alpha\beta \underline u}P_-+F_{(3)\alpha\beta\underline{\bar u}}P_+\right)\theta\label{eq:S8flux}\\
    &+\frac{\ep}{8\cdot3!}\overline\theta\Gamma^{\alpha\beta\gamma}\theta F_{(5)\alpha\beta\gamma\underline8\underline9}
    +\frac{i\ep}{8\cdot5!}\overline\theta\Gamma^{\alpha_1\cdots \alpha_5}\Gamma_{(8)}\theta F_{(5)\alpha_1\cdots \alpha_5}
    \bigg\}\,.\nonumber
\end{align}
In this action, we have used the holomorphic coordinate $u=x^8+ix^9$ such that \linebreak $F_{(3)\alpha\beta \underline u}\equiv (F_{(3)\alpha\beta \underline 8}-iF_{(3)\alpha\beta \underline 9})/2$ and $F_{(3)\alpha\beta\underline {\bar u}}\equiv (F_{(3)\alpha\beta \underline 8}+iF_{(3)\alpha\beta \underline 9})/2$ (and similarly for $H_{(3)}$). For later convenience we note that the last term can be rewritten as,
\be
    \frac{i\ep}{8\cdot5!}\overline\theta\Gamma^{\alpha_1\cdots \alpha_5}\Gamma_{(8)}\theta F_{(5)\alpha_1\cdots\alpha_5}
    \,=\,-\frac{\ep}{8\cdot3!}\,\overline\theta\Gamma^{\alpha\beta\gamma}\theta \,(\star_8F_{(5)})_{\alpha\beta\gamma}\,.
\ee

\subsection{Reduction to 6d}
\label{subsec:RedTo6D}

In (semi-)realistic type IIB models (or more generally in F-theory \cite{Weigand:2018rez}), chiral matter can be realized on the intersection locus of D7-brane stacks. The first step we perform is thus from 8d to the 6d fermionic fields localized on the intersection curve. In this subsection we derive the corresponding action following in particular \cite{Beasley:2008dc, Marchesano:2010bs}. We model the D7-branes intersection as follows: We view the CY locally as the product of three complex planes and assume that our $SU(M+N)$ stack fills out the first two factors:
$\mathcal D=\mathbb C_1\times\mathbb C_2 \subset 
\mathbb C_1\times\mathbb C_2\times \mathbb C_3$. We split the stack by turning on a scalar vev which is constant along $\mathbb C_1$ and proportional to the local coordinate $v$ parameterizing $\mathbb C_2$:
\begin{equation}
\vev \varphi = v\,m\,
\begin{pmatrix}N \one_{M}&\\&-M \one_N\end{pmatrix}\label{eq:phiVEV}\equiv v\, m\, \mathbf H\,.
\end{equation}
Here, $\mathbf H$ stands for the appropriately normalized Cartan generator and $m$ is a positive real number defining the angle between the intersecting stacks of $M$ and $N$ branes.

The breaking pattern is $SU(M+N)\rightarrow SU(M)\times SU(N)\times U(1)_{\mathbf H}$, with the adjoint representation decomposing as
\begin{equation}
\mathrm{adj}_{SU(M+N)}\rightarrow(\mathrm{adj}_{SU(M)},1)_0+(1,\mathrm{adj}_{SU(
N)})_0+(M,\overline N)_q+(\overline M,N)_{-q}+(1,1)_0\,.\label{eq:deco}
\end{equation}
At this point a subtlety arises: While the adjoint of $SU(M+N)$ is real, the two bifundamental representations are complex. This is in tension with our original spinor $\theta=\mathbf T^{a}\vartheta_{a}$ being Majorana and thus real. The resolution is to combine two of the $\vartheta_{a}$ in a complex spinor. In the particularly simple case $SU(2)\rightarrow U(1)$, one has
$\mathbf H\propto\sigma_3$ and $\theta\propto\sigma_i\vartheta_i$. It is then natural to define a generic (i.e. complex) 8d spinor $\vartheta\equiv\vartheta_1-i\vartheta_2$ and write
\be
\theta=\begin{pmatrix}\vartheta_{3}&\vartheta\\\mathcal B^{-1}\vartheta^*&-\vartheta_{3}\end{pmatrix},
\ee
where $\vartheta_{3}$ corresponds to the 
$U(1)_{\mathbf H}$ gaugino. The $SU(M+N)$ case is analogous,\footnote{In the case of F-theory, the same issue arises for exceptional groups. There, one has to combine states of opposing simple roots.}
\begin{equation}
\theta=\begin{pmatrix}\mathbf{T}^{a}\vartheta_{a}&\vartheta\\\mathcal B^{-1}\vartheta^\dagger
&\mathbf{T}^{b}\vartheta_{b}
\end{pmatrix}+\mathbf H\vartheta_{\mathbf H}\,,
\end{equation}
where $a=1,\dots,M^2-1$, $b=1,\dots,N^2-1$ and $\vartheta$ is the bifundamental. Note that the transposition implied in $\theta^\dagger$ acts only on the group indices, not on the spinor index. The subsequent discussion focuses only on $\vartheta$, as those are the states that will give rise to the SM matter.

The crucial bifundamental contribution in the action \eqref{eq:SYM8G} can be written as
\begin{equation}
    S_{\rm 8d}^{\rm YM}\supset\mu_{\rm D7}\!\int_{\mathcal D}\!\dd^8\xi e^{-\phi}\sqrt{-g}\,\tr\,\bigg\{\overline\vartheta \nabla_a\Gamma^a\vartheta+
    \begin{pmatrix}\overline\vartheta_+&\overline\vartheta_-\end{pmatrix}\underbrace{\begin{pmatrix}\nabla_{\alpha^\prime}\Gamma^{\alpha^\prime}&iq\overline\varphi\\
    -iq\varphi&\nabla_{\alpha^\prime}\Gamma^{\alpha^\prime}\end{pmatrix}}_{\textstyle  \equiv\,\,\tildeslashed{D}}
\begin{pmatrix}\vartheta_+\\\vartheta_-\end{pmatrix}\!\bigg\}\,,\label{eq:S8SYM2}
\end{equation}
with $\vartheta_\pm=P_\pm\vartheta$, $a=0,\dots,5$ and $\alpha^\prime=6,7$. Note that $\vartheta$ is $M\times N$ and thus the trace is now over an $N\times N$ matrix. For simplicity, we set the $U(1)$ charge $q$ to unity in what follows.

Given a set of 6d gamma matrices $\{\tilde\gamma^a\}_{a=0,\dots,5}$ and the chirality matrix $\tilde\gamma_{(6)}$, a good basis for the 8d to 6d reduction is
\begin{equation}
\Gamma^{\underline a}=\tilde\gamma^{\underline a}\otimes\one_2\,,\quad\Gamma^{\underline6,\underline7}=\tilde\gamma_{(6)}\otimes\sigma_{1,2}\,,\quad \Gamma_{(8)}=\tilde\gamma_{(6)}\otimes\sigma_3\,.\label{eq:Gamma8to6}
\end{equation}
Note that this basis is different from the purely imaginary $\Gamma^\alpha$'s which we used in section~\ref{man8d}. The advantage of this new basis choice is that both 6d and 8d chiralities are made nicely manifest. In particular, we have
\begin{align}
\vartheta_+&=\Psi^6_{+}\otimes\eta_{+}+\Psi^6_{-}\otimes\eta_{-}\,,\\
\vartheta_{-}&=\Psi^6_{+}\otimes\eta_{-}+\Psi^6_{-}\otimes\eta_{+}\,,
\end{align}
where $\Psi^6_\pm$ are 6d anticommuting spinors with $\tilde\gamma_{(6)}\Psi^6_\pm=\pm\Psi^6_\pm$ and $\eta_\pm$ are two 2d commuting spinors with $\sigma_3\eta_\pm=\pm\eta_\pm$.

From the action \eqref{eq:S8SYM2} we see that the operator $\tildeslashed{D}$ acts like a mass term for $\vartheta$. It gives rise to a single massless mode $\vartheta_0$, localized at the origin $v=0$ \cite{Marchesano:2010bs}. It is explicitly given by
\begin{equation}
\vartheta_0=\Psi^0_-\otimes\big(\eta_+-i\eta_-\big)\mathcal N\exp(- m|v|^2/2)\,,\label{eq:0Mode}
\end{equation}
where $\mathcal N$ is a normalization factor and, in our conventions, the mode has negative 6d chirality: $\tilde\gamma_{(6)}\Psi_-^0=-\Psi_-^0$. Substituting back into eq.~\eqref{eq:S8flux}, using \eqref{eq:Gamma8to6}, and integrating over $\mathbb C_2$ results in the 6d action (see appendix~\ref{app:Red} for details)
\begin{eqnarray}
S_{\rm 6d}&=&\mu\int \dd^6\xi\sqrt{-g}\,\,\tr\Bigg\{\,\overline\Psi_-^0\tilde\gamma^a\nabla_a\Psi_-^0
 \nonumber\\\label{eq:S6SUGRA} 
&&
+\frac1{12}\overline\Psi^0_-\tilde\gamma^{abc}\Psi^0_-\left[H_{(3)abc}+\frac\ep4 \left(F_{(5)abc\underline{89}}-(\star_8 F_{(5)})_{abc}\right)\right] \Bigg\}\\
&=& \mu\int \dd^6\xi\sqrt{-g}\,\,\tr\Bigg\{\,\overline\Psi_-^0\tilde\gamma^a\nabla_a\Psi_-^0+\frac1{12}\overline\Psi^0_-\tilde\gamma^{abc}\Psi^0_-\left[H_{(3)abc}+\frac\ep2F_{(5)abc\underline{89}}\right] \Bigg\}\,.\nonumber
\end{eqnarray}
Here, the normalization $\mathcal N$, $e^{-\phi}$ as well as factors arising from the integral over $\mathbb C_2$ have been absorbed in $\mu$. This quantity will not affect our results. The reason is that all operators we are interested in, i.e.~dipole operators and fermion kinetic terms, are bilinear in the fermions. Hence, $\mu$ can be changed arbitrarily by a fermion field rescaling. For convenience, we choose the dimension of $\mu$ consistently with the 4d and 2d spinors introduced in the next section having mass dimensions 3/2 and 1 respectively.

\section{Standard Model dipole coupling}
\label{sec:4d}

From the computation of the previous section it is not clear whether the couplings to the background will survive the mechanism generating 4d chiral matter: Indeed, looking at eq.~\eqref{eq:S6SUGRA} one would only expect vector-like pairs to interact with the SUGRA bulk fields. The SM, however, does not contain vector-like fermions.\footnote{Note though that the MSSM does.} As already mentioned in sec.~\ref{sec:VisSecCoup}, gauge invariance then requires that any dipole couplings emerging in the low-energy EFT involve the Higgs: 
\begin{equation}
X_{\mu\nu}\overline e H_d\sigma^{\mu\nu}\ell\,,\quad X_{\mu\nu}\overline Q H_u\sigma^{\mu\nu} u\,,\quad X_{\mu\nu}\overline Q H_d\sigma^{\mu\nu} d\,.
\label{eq:DipoleXSM}
\end{equation}
Hence, our strategy is to identify a toy-model brane construction in which the key chirality feature of the SM is realized. Based on this toy model, we will analyze whether the couplings in \eqref{eq:DipoleXSM} are really induced. Finally, we will estimate the suppression scale $\Lambda$ and the flavor structure of the dipole operators.

\subsection{The Pati--Salam toy model}
\begin{figure}
    \centering
    \includegraphics[scale=0.7
]{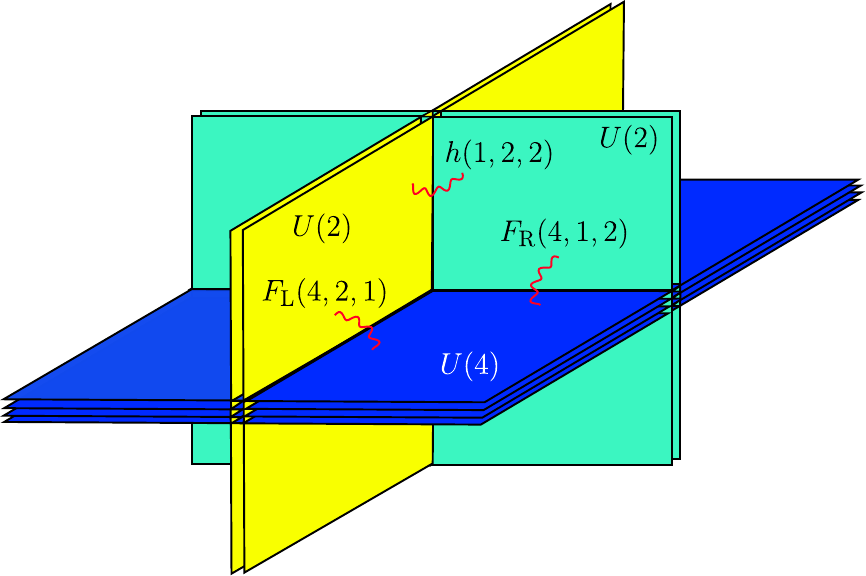}
    \caption{Local geometry of the realization of the Pati--Salam model. A $U(4)$ stack of branes and two $U(2)$ stacks intersect at right angles. The SM fermions live on the matter curves at the intersection of the $U(4)$ stack and the two $U(2)$ stacks while the Higgs live on the intersection of the two $U(2)$.}
    \label{fig:fig_ps}
\end{figure}
We consider a toy model that (locally) resembles the model studied in \cite{Cremades:2002qm,Cremades:2003qj,Lust:2004dn}. It is made out of three stacks of D7-branes, with gauge groups $U(4)$, $U(2)$ and $U(2)$ respectively, hence realizing the Pati--Salam setting \cite{Pati:1974yy}. Along the intersection two-cycles of two brane stacks, open strings generate bifundamentals, while flux along the $U(4)$ stack breaks the $\mathcal N=2$ structure, generating chiral matter (see fig.~\ref{fig:fig_ps}). This results in the classic Pati--Salam matter content, namely
\begin{equation}
\label{eq:PS}
F_{\rm L}(4,2,1)\oplus F_{\rm R}( 4,1,2)\oplus h(1,2,2)\rightarrow\big(Q\oplus \ell\big)\oplus\big(e\oplus u\oplus d\oplus \nu_{\rm R}\big)\oplus\big(H_d\oplus H_u\big)\,.
\end{equation}
By choosing three units of flux on the $U(4)$ stack one thus exactly reproduces the MSSM particle content plus right-handed neutrinos.

Important for the purpose of this paper is that this model has Yukawa interactions of the type
\begin{equation}
Y_{ij} \overline e^i H_d\ell ^j\,,\quad Y_{ij}\overline u^i H_u Q^j\quad\text{ and }\quad Y_{ij}\overline d^i H_d Q^j\,,\label{eq:PSYukawa}
\end{equation}
capturing the chiral structure and the Higgs mechanism of the SM. They are generated at (complex) co-dimension three loci where all the three stacks meet and can be computed explicitly, either by a string worldsheet computation (see \cite{Cvetic:2003ch,Abel:2003yx,Lust:2004cx} in type IIB or \cite{Cremades:2003qj} for the T-dual IIA computation), or with field theory techniques \cite{Cremades:2004wa}. This work will be concerned with the latter as it allows to directly make contact with the previous sections. In the case at hand
\begin{equation}
Y_{ij}\propto \zeta_{F_{\rm L}^i}(z_c,\overline z_c)\zeta_{F_{\rm R}^j}(z_c,\overline z_c)\zeta_h(z_c,\overline z_c)\,,\label{eq:Y-wave}
\end{equation}
where $\zeta_i$ are the matter wavefunctions in the presence of flux along the matter curves $\Sigma_i$ evaluated at the intersection point $z_c$ where all curves meet \cite{Cremades:2004wa}. For certain choices of $\Sigma_i$, the $\zeta_i$ have been explicitly computed (for tori see e.g. \cite{Cremades:2004wa}, or \cite{Conlon:2008qi} for simple projective spaces). Note that \eqref{eq:Y-wave} implies that, at leading order, $Y_{ij}$ is of rank one, similarly to Yukawas originating from co-dimension three singularities in the context of F-theory.

\subsection{4d effective theory from a generic matter curve}\label{mat_curv_to_4d}

Adapting the 6d action of \eqref{eq:S6SUGRA} to the toy model of the previous subsection, we can now derive the 4d EFT following, e.g.,~\cite{Cremades:2004wa,Buchmuller:2016gib}. The procedure is similar to the 8d-to-6d reduction of sect.~\ref{subsec:RedTo6D}. While one could also go directly from 8d to 4d as in~\cite{Marchesano:2010bs, Conlon:2008qi}, without the intermediate 6d EFT, we feel that our procedure of compactifying a fluxed 6d theory makes the genericity of our results more transparent.

Note that, when compactifying a 6d charged hypermultiplet to 4d on a Riemann surface with $F_{(2)}$-flux, SUSY breaks completely. However, in the type of string models we are interested in, 4d ${\cal N}=1$ SUSY is known to survive. This is achieved by ensuring that the $F_{(2)}$-flux on the complex surface wrapped by one of the two underlying D7-brane stacks is supersymmetric. In turn, this implies non-zero flux components transverse to the 6d surface, which one would have to include in the 6d EFT in the form of complex scalar fields with appropriate profiles and couplings to 6d matter. We disregard this interesting aspect, focusing exclusively on the charged fermions and their resulting 4d couplings. We will thus not be able to see 4d SUSY emerge. However, this shortcoming affects our results only by ${\cal O}(1)$-factors, which are not important since we anyway do not employ an explicit and realistic background geometry and brane content.

Our goal is to reduce an action involving the 6d Dirac operator $\slashed \nabla_6$ to 4 dimensions. We assume a generic intersection geometry and the specific case of a flat torus is developed in appendix \ref{app:MagTor}. To be explicit, we choose 6d gamma matrices
\begin{equation}
    \tilde\gamma^\mu=\gamma^\mu \otimes\one_2\,,\quad\tilde\gamma^{\underline4,\underline5}=\gamma_{(4)}\otimes\sigma_{1,2}\,,\quad \tilde\gamma_{(6)}=\gamma_{(4)}\otimes\sigma_3\,,\label{eq:Gamma6to4}
\end{equation}
such that
\begin{equation}
\slashed\nabla_6=\slashed\nabla_4\otimes\one+\gamma_{(4)}\otimes\underbrace{\begin{pmatrix} 0 & \nabla_z\\
\nabla_{\bar z} & 0 \end{pmatrix}}_{\textstyle  \equiv\,\,\slashed\nabla_2}\,.
\end{equation}
Here, the 2d covariant derivative $\nabla_z$ involves the brane flux.

Our 6d fermion is chiral and may hence be factorized symbolically as
\begin{equation}
\Psi_-^0\,\,\sim\,\,\sum \Big( \psi_+(x)\otimes\chi_-(z)+\psi_-(x)\otimes\chi_+(z)\Big) \,,
\end{equation}
where $\gamma_{(4)}\psi(x)_\pm=\pm\psi(x)$ and $\sigma_3\chi_\pm(z)=\pm\chi(z)$.
Given that on a smooth, compact Riemannian manifold, the Dirac operator $\slashed{\nabla}_2$ is essentially self-adjoint, it appears natural to choose the internal wavefunctions as its eigenmodes. This, however, is impossible because we need our modes to be chiral while the Dirac operator changes chirality. As is well established \cite{Green:1987mn}, one proceeds by using $(\slashed{\nabla}_2)^2$. The latter commutes with the internal chirality matrix and can hence be diagonalized in an orthonormal, chiral basis,
\begin{equation}
(\slashed{\nabla}_2)^2\chi_{\pm n}(z)=m_n^2\chi_{\pm n}(z)\,.
\end{equation}
For any $\chi_{\pm n}(z)$, the state $\slashed{\nabla}_2\chi_{\pm n}(z)$ is also an eigenstate at the same level, such that massive eigenstates always come in pairs of opposite chirality,
\begin{equation}
\label{eq:decomposition}
\slashed{\nabla}_2\chi_{\pm n}(z)
=m_n\chi_{\mp n}\,.
\end{equation}
No such pairing arises at the zero-mode level. Instead, due to the index theorem, a number of chiral zero modes corresponding to the amount of gauge flux are present. Without loss of generality we choose the sign of the flux such that these modes are $\chi_{-0}^i(z)$, being also orthonormal and labeled by the index $i$. We ignore the possible presence of massless vector-like pairs for now but comment on this at the end of sect.~\ref{d4dd}.

We may now be precise and decompose our 6d chiral field as
\begin{equation}
\Psi_-^0=\sum_i\psi_{+0}^i(x)\otimes\chi_{-0}^i(z)+\sum_{n>0}\Big(\psi_{+n}(x)\otimes\chi_{-n}(z)+\psi_{-n}(x)\otimes\chi_{+n}(z)\Big)\,.
\end{equation}
The 6d kinetic term becomes\footnote{The $\psi$ and $\chi$ fields have mass dimension 3/2 and 1 respectively, cf.~discussion below~\eqref{eq:S6SUGRA}.}
\begin{align}
&\overline\Psi_-^0\slashed\nabla_6\Psi_-^0
=
\sum_{i,j}\left[\overline\psi_{+0}^i(x)\slashed\nabla_4\psi_{+0}^j(x)\right]\left[(\chi_{-0}^i)^\dagger(z)\chi_{-0}^j(z)\right]\\
&+\sum_{n,m>0}\Bigg\{\left[\overline\psi_{\pm n}(x)\slashed\nabla_4\psi_{\pm m}(x)\right]\left[\chi_{\mp n}^\dagger(z)\chi_{\mp m}(z)\right]
\mp\left[\overline\psi_{\pm n}(x)\psi_{\mp m}(x)\right]\left[\chi_{\mp n}^\dagger(z)\slashed\nabla_2\chi_{\pm m}(z)\right]\Bigg\}\,.\nonumber
\end{align}
Here, an appropriate summation over terms with upper and lower signs ($\pm$) is understood on the r.h.~side.
Using \eqref{eq:decomposition} and the orthonormality relations of the 2d modes, the integration over the internal space yields the free 4d Lagrangian
\begin{align}
\label{eq:reduc_kinetic}
&\sum_i\overline\psi_{+0}^i(x)\slashed\nabla_4\psi_{+0}^i(x)+\sum_{n>0}\bigg\{\overline\psi_{\pm n}(x)\slashed\nabla_4\psi_{\pm n}(x)\pm m_n\overline\psi_{\mp n}(x)\psi_{\pm n}(x)\bigg\}\,.
\end{align}

In addition to the kinetic term, the action \eqref{eq:S6SUGRA} contains the coupling of the 6d spinor to the bulk fluxes. It involves either $H_{(3)}$, $F_{(5)}$ or its dual and may be schematically written as $\overline\Psi_-^0\slashed{\mathcal A}\Psi_-^0$, where $\mathcal A$ is a 3-form. Because we are interested in generating 4d dipole couplings of the form \eqref{eq:DipoleXSM}, we may restrict attention to flux terms where strictly two indices are along the 4d spacetime. The relevant gamma-matrix structure then reads
\begin{equation}
\gamma^{\mu\nu}\gamma_{(4)}\otimes\begin{pmatrix}
    0 & \mathcal A_{\mu\nu \underline z}(z)\\ \mathcal A_{\mu\nu \overline {\underline z }}(z) & 0
\end{pmatrix}.
\end{equation}
As we will see, after KK reduction to 4d our interest is only in insertions of this term between a zero mode and a massive mode. The corresponding part of the 4d Lagrangian reads
\begin{equation}
\sim\mu\!\sum_{n>0,\,i}\!\overline\psi_{-n}(x)\gamma^{\mu\nu}\psi_{+0}^i(x)\!\int\dd z\dd\bar z\,\sqrt{h}\left[\chi_{+n}^\dagger(z)\!\begin{pmatrix}
    0 & \mathcal A_{\mu\nu \underline z}(z)\\ \mathcal A_{\mu\nu \overline {\underline z}}(z) & 0
\end{pmatrix}\!\chi_{-0}^i(z)\right]+\text{h.c.}
\label{ani}
\end{equation}
where $h$ denotes the determinant of the 2d internal metric. At this point we have to remember that what was so far treated as a 3-form or 5-form background flux is actually the field strength profile associated with a dynamical 4d massless vector. This vector is either a hidden-brane or RR photon. Calling its 4d field strength $X_{\mu\nu}$ we may then substitute 
\be
{\cal A}_{\mu\nu \underline z}(z)\,\,\,\to\,\,\,
X_{\mu\nu}(x)\,f(z)\,\,,\qquad \qquad {\cal A}_{\mu\nu \overline z}(z)\,\,\,\to\,\,\,
X_{\mu\nu}(x)\,\overline f(z)\,.\label{eq:Asub}
\ee
Here, the function $f$ encodes the appropriate internal profile. Inserting this in \eqref{ani} and defining coefficients $a_n^{i}$ as integrals over $\chi_{-n}^*$, $\chi_{+0}^i$ and $f$ we arrive at
\begin{equation}
\frac{1}{\Ms}\sum_{n>0,\, i} a_n^iX_{\mu\nu}\overline\psi_{-n}(x)\gamma^{\mu\nu}\psi_{+0}^i(x)+\text{h.c.}\,,
\label{apsi2}
\end{equation}
where we introduced the string scale $M_{\rm s}=1/\sqrt{\alpha'}$. In what follows we want to be more precise about the scaling of $a_n^i$ with radii and $g_s$ for the brane and RR photons.

\paragraph{Brane photons: } From what has been said above, we explicitly have 
\begin{equation}
a_n^i=\Ms\int\dd z\dd\bar z\,\sqrt{h}\,\chi_{+n}^\dagger(z)\begin{pmatrix}0 & f(z)\\ \bar f(z) & 0\end{pmatrix}\chi_{-0}^i(z)\,.
\end{equation}
We assume that the D3-brane sits at a distance $L$ from the matter curve and we neglect the variation of the flux in the domain of integration, i.e.~$f(z)\simeq\,\mbox{const.}$ The value of this constant can be read off from the $F_{(3)}$ and $H_{(3)}$ profiles in \eqref{eq:F3H3}, with $r$ replaced by $L$. One also has to take into account the numerical factor arising from the replacement of $F_{(2)\mu\nu}$ with the canonically normalized field strength $X_{\mu\nu}$. Given that the $F_{(2)}$ kinetic term in the DBI action \eqref{eq:DBI} comes with a factor $1/(8\pi\gs)$, one finds 
$F_{(2)\mu\nu}=X_{\mu\nu}\sqrt{2\pi\gs}$. Finally, the numerical factor that we omitted in \eqref{ani} can be read from \eqref{eq:S6SUGRA}: For brane photons the factor $1/12$ is multiplied by a combinatorial factor $3$ to get the correct index structure, and an additional factor 2 to go to a complex third index. Combining all of this gives
\begin{equation}
a_n^i=2^\frac{9}{2}\pi^\frac{5}{2}\gs^\frac{3}{2}\left(\frac{\sqrt{\alpha'}}{L}\right)^5\underbrace{\int\dd z\dd\bar z\,\sqrt{h}\, \chi_{+n}^\dagger(z)\begin{pmatrix}0 & 1\\ 1 & 0\end{pmatrix}\chi_{-0}^i(z)}_{\approx\, 1\,\,\,\text{for low}\,\,\,n}\,.\label{eq:ap_int}
\end{equation}

\paragraph{RR photons: } The case of the RR photon is analogous but slightly more involved. Looking at \eqref{eq:F5self} one naively expects two contributions to the substitution \eqref{eq:Asub}: One originating from $\mathsf F^\kappa_{(2)}$ and one from $\star_4\mathsf F^\kappa_{(2)}$. However, the second term can be traded for an additional contribution proportional to $\mathsf F_{(2)}^\kappa$ by inserting a $\gamma_{(4)}$ in front of $\psi_{+0}^i$ in equation \eqref{ani} and then applying the identity \eqref{eq:gamma5tostar}. This results in 
\begin{equation}
 {\cal A}_{\mu\nu \underline z}(z)\,\,\,\to\,\,\,
 \mathsf X^\sigma_{\mu\nu}(x)\,f_\sigma(z)\,,
\end{equation}
with
\begin{equation}
f_\sigma(z)=\sqrt{2}\kappa\,{\tilde C_{\sigma}}\,^{\kappa}\Big[(\alpha_\kappa)_{\underline{z89}}+\big((i\one -A)C^{-1}\big)_{\kappa\lambda}(\beta^\lambda)_{\underline{z89}}\Big]\,,\label{eq:fRR}
\end{equation}
and where the matrices $A$ and $C$ have been defined in \eqref{eq:abstar}, while $\tilde C$ is a field redefinition $\mathsf X^\sigma{\tilde C_{\sigma}}\,^{\kappa}=\mathsf F^\kappa_{(2)}$ which, together with the factor $\sqrt2\kappa$, canonically normalizes the kinetic term in eq.~\eqref{eq:RR4d}. In a generic geometry of two intersecting D7-branes, D7$_{\rm a}$ and $D7_{\rm b}$, the meaning of the indices $\underline{89}$ is not immediately obvious: Those contractions arose in section \ref{man8d} in the limit where the two intersecting banes have only a small relative angle, characterized by a sufficiently flat profile of the vev of $\varphi$. The directions $\underline 8$ and $\underline 9$ are then unambiguously defined as being orthogonal to the original brane stack with $\varphi\equiv 0$. Branes intersecting at large angles are then formally treated by giving the vev $\varphi$ a steep profile near the intersection locus $\phi=0$. This may not provide the exact result but it is sufficient at our level of precision.

We see that the three-forms $\alpha_\kappa,\,\beta^\kappa$ appear in \eqref{eq:fRR} with flat indices, i.e.~they secretly contain three inverse vielbeins. This implies that the coefficients $a_n^i$ derived from this behave as $a_n^i\propto 1/\sqrt{\vol}$ when the CY is scaled isotropically. This matches with the findings of \cite{Anastasopoulos:2020xgu}. However, this naive scaling does not necessarily reflect the complete picture. Indeed, we now focus on the favourable case where a relevant three-cycle is much smaller than the square root of the CY volume $\sqrt{\cal V}$. This happens, for example, near the conifold degeneration of a CY. In the conifold case, it is easy to convince onself that the flat-index coefficients of the two harmonic three-forms, $\alpha_{\underline{z89}}$ and $\beta_{\underline{z89}}$, scale with $r^{-3}$, where $r$ is the radial coordinate and the deformed region corresponds to $r_{\rm min}=R$. This behaviour implies that the integrals of $|\alpha|^2$ and $|\beta|^2$ localize near the deformed region, up to a logarithmic UV tail which we may ignore. The only relevant scale is then the scale $R$ of the deformation, i.e.~the scale of the small 3-cycle. Thus, if the D7s realizing the SM are in the vicinity of such a small, localized three-cycle and the volume grows large (e.g. by making one or more four-cycles big) the local geometry relevant for the calculation of the $a_n^i$ remains completely determined by the scale $R\ll {\cal V}^{1/6}$. We then expect $a_n^i\sim 1/R^3$, without volume suppression.\footnote{In the extreme case this can  lead to light charged states for the RR photons. For example as one shrinks the $S^3$ of the conifold to zero size the mass of a D3 wrapping the $S^3$ goes to zero as well, such that our RR photon is not so superhidden anymore. We thank Jakob Moritz for pointing this out to us.}.

To estimate the prefactors in this best-case scenario, we use a rectangular $T^6$ toy model with radius $R$. Of course, this geometry can not be ``glued'' to a large CY with ${\cal V}\gg R^6$, but our goal here is simply to make sure that we do not overlook some conceivable accumulation of $(2\pi)$ factors, as it indeed appears in the brane photon case. The details of our very simple analysis are given in appendix \ref{app:RRTor}. The equivalent of \eqref{eq:ap_int} is then found to be
\begin{equation}
a_n^i=\sqrt{\frac{\pi}{2}}{\frac{\gs}{4}}\frac{(\ap)^{3/2}}{R^3}\underbrace{\int\dd z\dd\bar z\,\sqrt{h}\,\chi_{+n}^\dagger(z)\begin{pmatrix}0 & 1\\ 1 & 0\end{pmatrix}\chi_{-0}^i(z)}_{\approx\, 1\,\,\,\text{for low}\,\,\,n}\,.\label{eq:ap_intRR}
\end{equation}
Recall that this formula is an estimate for the favourable case with small RR-photon three-cycle and small D7-brane four cycle of radius $R\ll{\cal V}^{1/6}$. We leave the discussion of other relevant situations, e.g.~with a three-cycle much larger than the D7 but much smaller than typical CY scales, for the future.

\subsection{Deriving the 4d dipole operator}
\label{d4dd}

In the last subsection, we have derived the building blocks in the 4d EFT which are needed to obtain the desired dipole operator \eqref{eq:DipoleXSM}. To spell this out, we recall that our generic matter curve of the last subsection stands for any of the 3 curves intersecting at a point responsible for generating a SM Yukawa interaction, as in the example of~fig.~\ref{fig:fig_ps} and \eqref{eq:PS}. To be concrete, let us assume $\psi_{+0}^i$ of the last subsection stands for the lepton-doublet zero modes, with $i\in\{1,2,3\}$ labeling the family or flavor. The other two matter curves of fig.~\ref{fig:fig_ps} then supply the Higgs doublet $H$ and the right-handed-fermions (electron, muon or tau) $e^i$, see again \eqref{eq:PS}.

\begin{figure}[ht]
    \centering
    \includegraphics[scale=.9]{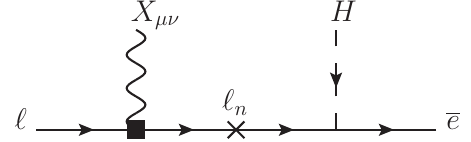}
    \caption{Feynman diagram calculating the contribution to the dipole operator 
    $X_{\mu\nu}\overline e H_d\sigma^{\mu\nu}\ell$ that arises from integrating out massive KK modes of the lepton field $\ell$.}
    \label{fig:mediationTree}
\end{figure}

The desired 4d dipole operator now arises after integrating out the tower of KK modes of the lepton-doublet $\ell$. The corresponding Feynman diagram, displayed in fig.~\ref{fig:mediationTree}, is based on three crucial contributions. Following the diagram from left to right these are:

\begin{itemize}
\item The coupling of type \eqref{apsi2} between the 4d lepton zero modes $\ell^i$, the superhidden photon and higher KK modes $\ell_n$ of the 6d field associated with the lepton:
\be
\frac{1}{\Ms}\sum_i\,a_n^i X_{\mu\nu}\overline\ell_n\gamma^{\mu\nu}\ell^i\,.
\ee
\item The mass terms for these KK mode as given in \eqref{eq:reduc_kinetic}:\qquad $m_n\overline\ell_n\ell_n$.
\item The Yukawa interactions between the massive KK modes $\ell_n$, the right-handed lepton zero modes $e^i$, and the Higgs zero mode. These are analogues of the SM Yukawa couplings between the three chiral massless modes coming from the three curves, known to be proportional to the values of the mode profiles at the interaction point. We call the coefficients $y_n^{i}$ such that the operator reads \qquad $\displaystyle\sum_i y_n^i\overline e^iH_d \ell_n \,\,.$
\end{itemize}

The diagram then contributes to the following dipole term:
\begin{equation}
\mathcal L_{\rm dipole}=\sum_{i,j}\frac{c_{ij}}{ \sqrt{2}\Lambda^2}X_{\mu\nu}\overline e^i H_d\gamma^{\mu\nu}\ell^j{ +\text{ h.c.}}\label{eq:resXcoup}
\end{equation}
Here we introduced a factor $\sqrt{2}$ in the denominator to match the normalization of \eqref{eq:dipoleint}, taking into account that in our conventions the Higgs field vev will contribute a factor $v_{\rm h}/\sqrt{2}$. The combination $c_{ij}/\Lambda$ is now unambiguously defined and given by the expression
\begin{equation}
\frac{c_{ij}}{\sqrt{2}\Lambda^2}=\frac{1}{\Ms}\sum_{n>0}\bigg(\frac{y_n^{i}a_n^j}{m_{n}}+\frac{\tilde y_n^{i}\tilde a_n^j}{\tilde m_{n}}\bigg)\,.
\label{coms}
\end{equation}
Here, the second term with the tilde notation accounts for a process analogous to that of fig. \ref{fig:mediationTree} but with the matter curves corresponding to $\ell$ and $e$ exchanged. In other words, $\tilde{m}_n$ are masses of the right-handed lepton KK modes, $\tilde{y}_n^i$ are the corresponding Yukawa interaction coefficients and $\tilde{a}_n^j$ are the $X^{\mu\nu}$ couplings. We estimate the scale $m_1$ to be the typical mass scale of the SM-field KK towers, without taking into account corrections from fluxes: $m_1\equiv 1/R_{\rm D7}$. For definiteness, we may take $2(R_{\rm D7}/\sqrt{\alpha'})^4\gs^{-1}\simeq \alpha_{\rm SM}^{-1}\simeq 30$, given that SM gauge couplings are typically of that size and are set by the D7-stack volumes.\footnote{
One 
may check that explicitly $m_1\simeq \Ms/(2g_s^{1/4})$ with, in our conventions, the lightest open-string excitations sitting at $\Ms$.
} This expression simply states that the inverse gauge coupling is given by two times the Einstein frame volume.
For simplicity we take $g_{\rm s}=1$, but it would be easy to reinstall this parameter.

We now proceed by treating \eqref{coms} as follows. We define $\sqrt{2}\Lambda^2$ as $\Ms m_1$ divided by twice the prefactor coming with $a_n^i$ (to take into account the tilde contribution). This prefactor is specified in \eqref{eq:ap_int} for brane photons and in \eqref{eq:ap_intRR} for RR photons, where we also set $R=R_{\rm D7}$. The leading ($n=1$) contribution to $c_{ij}$ is then determined by $y_1^i$, $\tilde{y}_1^i$, and the ${\cal O}(1)$ integrals associated with $a_1^i$, $\tilde{a}_1^i$. This allows us to formulate a key preliminary result:

\begin{framed}
\noindent
Our prediction for the dipole operator is given by \eqref{eq:resXcoup} with
\begin{align}
&\text{\underline{Brane photons:}}\qquad &&\Lambda^2\equiv\frac{\Ms^2}{128\,\pi^{5/2}}\,\left(\frac{L}{\sqrt{\alpha'}}\right)^5\,,\label{eq:exp_Lambda_brane}\\
&\text{\underline{RR photons:}}\qquad &&\Lambda^2\equiv\frac{8}{\sqrt{\pi}}\Ms^2\,,
\label{eq:exp_Lambda_RR}
\end{align}
and with coefficients $c_{ij}$ that are ${\cal O}(1)$ in the absence of flavor suppression. Flavor will be discussed in sect.~\ref{sec:yuk}. There, we will see that our analysis concretely implies that the third-generation entry is $c_{33}\sim 1$ with the above definition of $\Lambda$ and other entries may or may not be smaller in a model-dependent way.
\end{framed}

Several remarks are in order. First, note that for brane photons $\Lambda$ enjoys a significant suppression by powers of `$\pi$' and `2'. Next, the convergence of the sum in \eqref{coms} is obviously crucial: Recall that the coefficients $a_n^i$ arise as 2d integrals of a product of the bulk $X^{\mu\nu}$ profile pulled back to the matter curve, the zero mode $\ell^i$ and one of the corresponding higher modes $\ell_n$. The tower of these higher modes forms an orthonormal basis on the curve -- the flux-modified analogue of the Fourier basis. The integral of these modes with the smooth function formed by the $\ell$ and the $X^{\mu\nu}$ profile is hence expected to fall off exponentially fast with growing $n$. Thus, the sum is expected to converge fast. This can be confirmed by the case of the flat torus analyzed explicitly in appendix~\ref{app:MagTor}. In fact, for constant $X^{\mu\nu}$ profile one can see that only a finite number of the $a_n^i$ are non-zero.
 
\begin{figure}
    \centering
    \includegraphics[scale=.9]{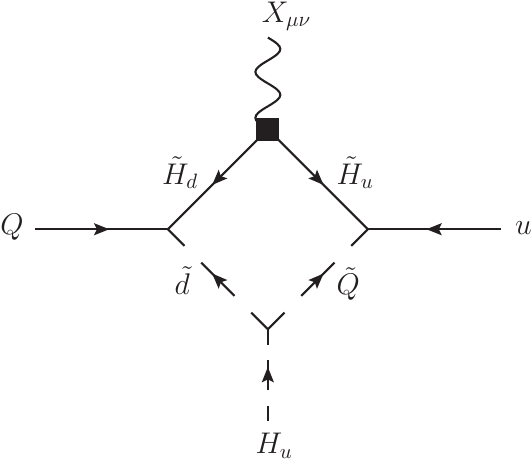}
    \caption{Example of a process mediated by SUSY partners, provided the existence of the $\overline{\tilde H_u}\gamma^{\mu\nu}\tilde H_d X_{\mu\nu}$ operator. $\tilde H_u$ and $\tilde H_d$ are the superpartners of $H_u$ and $H_d$ while $\tilde d/\tilde Q$ denote the down/doublet squarks. The arrows in the diagram are drawn according to the two-component formalism often used in SUSY phenomenology.}
    \label{fig:mediationLoop}
\end{figure}

A further comment concerns competing loop effects. So far, we have been agnostic concerning the SUSY breaking scale. We do not want to commit to a specific SUSY breaking source and thus $m_{\rm SUSY}$ could be anywhere between $\Ms$ and the LHC lower bounds. In particular, SUSY partners can be lighter than $m_{\rm KK}$. Similar to dimension-five proton decay in SUSY GUTs, it is conceivable that instead of the tree-level process of fig.~\ref{fig:mediationTree}, operators such as those in \eqref{eq:DipoleXSM} are generated via loops involving SUSY partners instead of KK modes, see fig.~\ref{fig:mediationLoop}. Denoting the higgsinos by $\tilde H_d,\,\tilde H_{u}$, the MSSM in principle allows for the appearance of a term like $\overbar{\tilde H_u}\gamma^{\mu\nu}\tilde H_d X_{\mu\nu}$, hence higgsinos and squarks\footnote{The absence of a right-handed neutrino in the MSSM forbids a similar process for sleptons} could run in the loop. We expect the crucial prefactor in \eqref{eq:exp_Lambda_brane} and \eqref{eq:exp_Lambda_RR} to be replaced according to 
\begin{equation}
M_{\rm s}^2\qquad \longrightarrow\qquad M_{\rm s} m_{\rm SUSY}/c_{\rm loop}\,,
\end{equation}
where $c_{\rm loop}$ encodes the loop suppression. Note that in the particular example of fig.~\ref{fig:mediationLoop} there is also an additional suppression originating from the smallness of the Yukawas.

Last but not least, it is also conceivable that the compactification gives rise to vector-like particles with the same quantum numbers as the states in the KK towers but with masses $m_{\rm VL}\ll m_{\rm KK}$. This again would allow for processes as depicted in fig.~\ref{fig:mediationTree} and would lead to an enhancement in \eqref{eq:exp_Lambda_brane} and \eqref{eq:exp_Lambda_RR}:
\begin{equation}
M_{\rm s}^2\qquad \longrightarrow\qquad M_{\rm s} m_{\rm VL}\,.
\end{equation}

In particular, in models with (at least two) r.h.~neutrinos, those could play the role of the vector-like states above. A diagram analogous to that in fig.~\ref{fig:mediationTree}, but with two Higgs vertices left and right of the $X_{\mu\nu}$ vertex, then induces an operator of the form
\be
\mathcal L_{\rm dipole}\supsetsim \frac{y_\nu^2}{\Ms}\Big(\frac {v_{\rm h}}{M_\nu}\Big)^2(\overline\nu^i)^c\gamma^{\mu\nu}\nu^
jX_{\mu\nu}\,,
\ee
with $M_\nu$ the mass of the r.h.~neutrino and $y_\nu$ its Yukawa coupling. Given the constraint on the physical neutrino mass scale, $m_\nu\sim y_\nu^2 v_{\rm h}^2/M_\nu \lesssim 1$\,eV (see e.g.~\cite{KATRIN:2024cdt}), the resulting dipole can be relevant if one is prepared to assume low values of $M_\nu$. Specifically, translating to the parametrization of \eqref{eq:dipole_intro} gives $\Lambda \sim (M_\nu/y_\nu)\sqrt{\Ms/v_{\rm h}}\gtrsim 5\times 10^5 \sqrt{\Ms M_\nu}$. For example, for $M_\nu\sim 1$\,keV and $\Ms\sim 10^3$\,TeV, the effective suppression scale becomes $\Lambda\sim 5\times 10^5\,$GeV. 

Of course, integrating out such light r.h.~neutrinos is not appropriate if the energy scale of a given experiment or observation is above $M_\nu$. One then has to resum the Higgs-neutrino-neutrino vertex to all orders, leading to the parametrically very similar result
\be
\mathcal L_{\rm dipole}\supsetsim\frac {\theta_{i}\theta_j}\Ms(\overline \nu^i)^c\gamma^{\mu\nu}\nu^jX_{\mu\nu}\,.
\ee
Here $\theta_i\sim\sqrt{m_\nu/M_\nu}$ are mixing angles between active and sterile neutrinos. Comparing again with our previous parameterization, we have 
$\Lambda\simeq\sqrt{\Ms v_{\rm h}}/\theta$, which matches our previous estimate for appropriate choices of $m_\nu$, $M_\nu$ and $\Ms$. It can of course become even lower in the regime of very small $M_\nu$ and large mixing angle. We leave a more careful phenomenological analysis to future work.

\subsection{Flavor structure}
\label{sec:yuk}

We now analyze the flavor structure of the dipole operator, as encoded in the coefficients $c_{ij}$. While such a set of coefficients exists for up-type quarks, down-type quarks and leptons, we will, as before, primarily focus on the leptonic sector for definiteness. Indeed, the same mechanisms that give rise to the flavor hierarchies of SM fermion masses may also leave characteristic imprints on the flavor structure of $c_{ij}$. In order to illustrate this, we will consider the following three possibilities:
\begin{enumerate}
\item The structure in the Yukawa interactions is \textit{anarchic}. In other words, there is {\it no} particular mechanism behind the flavor structure of SM fermion masses.
\item At leading order the Yukawa matrix is of rank one, making the third family heavy. Second and first generation masses are created by a subleading correction of the {\it same} size, at least parametrically.
\item At leading order the Yukawa matrix is of rank one. Second and first generation masses come from distinct effects of {\it successively weaker} strength.
\end{enumerate}
All of the above have a natural stringy realization in terms of the precise geometry and intersection pattern of the matter curves. What is common to all scenarios is that the three generations of left-handed leptons $\ell$ come from one curve and the three generations of right-handed leptons $e$ come from another curve. In addition, there is also the down-type Higgs curve.

The first, anarchic case is obtained if these three curves intersect at (at least) three different points in the CY space. From~\eqref{eq:Y-wave} it is obvious that each intersection contributes a rank-one matrix with $\mathcal O(1)$ elements to the total Yukawa coupling. The SM fermion mass hierarchy is then purely accidental. Since the computation of $c_{ij}$ involves integrating bulk $p$-form field profiles and zero-mode wavefunctions along the curves (see \eqref{eq:ap_int}), each intersection point contributes very differently to $c_{ij}$ than to $Y_{ij}$. We then expect that, after adding the contributions of all points and going to a basis where $Y_{ij}$ is diagonal, the $c_{ij}$ will form a random $3\times 3$ matrix with ${\cal O}(1)$ coefficients.

The other two cases arise naturally when the three relevant matter curves intersect only once. This single intersection then gives, at leading order, a rank-one Yukawa matrix. Since all generations live on the same curves, it is always possible to perform a redefinition in flavor space such that only one flavor has non-vanishing wavefunction at the intersection point. The resulting Yukawa can schematically be written as
\begin{equation}
Y_{ij}\sim\begin{pmatrix}0\\0\\ *\end{pmatrix}\big(0\quad0\quad*\big)=\begin{pmatrix}0&0&0\\0&0&0\\0&0&*\end{pmatrix}.\label{eq:yuk.esti}
\end{equation}
Assuming that \eqref{eq:Y-wave} also holds for the Yukawas of the higher KK modes $y^i_n,\,\tilde y^i_n$, one has
\begin{equation}
y^i_n,\,\tilde y^i_n\sim\big(0\quad0\quad *\big)\,.\label{eq:yuk.n}
\end{equation}
Inserting this in equation \eqref{coms} yields, for the single intersection and at leading order,
\begin{equation}
c_{ij}\sim\begin{pmatrix}0&0&*\\0&0&*\\ *&*& *\end{pmatrix}\,.\label{eq:c.esti}
\end{equation}
Here, we assumed generic $a^i_n,\,\tilde a^i_n$. 

By adding subleading corrections to $Y_{ij}$ in \eqref{eq:yuk.esti}, re-diagonalizing with a bi-unitary transformation, and then applying the transformation to \eqref{eq:c.esti}, one can estimate the corresponding effect on $c_{ij}$. For example, one may assume that the electron and muon masses are generated by a correction of the form $Y\rightarrow Y+\varepsilon Y_{\mu e}$ where $\varepsilon\sim m_\mu/m_\tau$ and $Y_{\mu e}$ is a full rank matrix with order one entries. This corresponds to our `case 2', where the lightness of the electron remains accidental. The resulting $c_{ij}$ has the form
\be
c_{ij}\sim\begin{pmatrix}\varepsilon&\varepsilon&*\\\varepsilon&\varepsilon&*\\ *&*&*\end{pmatrix}.\label{eq:flavour}
\ee

Finally, in our `case 3', fermion masses are generated by successive rank-one corrections $Y\rightarrow Y+\varepsilon Y_{\mu}+\varepsilon^\prime Y_e$, where $\varepsilon^\prime\sim m_e/m_\tau$. One then naively expects
\be
\label{eq:flavour2}
c_{ij}\sim\begin{pmatrix}\varepsilon^\prime&\varepsilon&*\\\varepsilon&\varepsilon&*\\ *&*&*\end{pmatrix}.
\ee
However, it is conceivable that \eqref{eq:flavour} is in fact the general case as one could imagine that the rank-one correction generating $m_\mu$ also changes the KK Yukawas to  $y^i_n\sim(\varepsilon\quad\varepsilon \quad1)$.

\bigskip 

We close this section by providing, for convenience, a dictionary between the couplings used above and the more phenomenological notation \eqref{eq:dipoleint} for the dipole operator from the introduction. This will be heavily used in the next section. The dictionary is:
\begin{align}
\begin{split}
\label{eq:ReCImC}
&d_{ij}^{\rm M}=\frac{i}{2}\big(c^\dagger-c\big)_{ij}\,,\\
&d_{ij}^{\rm E}=-\frac{1}{2}\big(c+c^\dagger\big)_{ij}\,.
\end{split}
\end{align}

\section{Phenomenological constraints}
\label{sec:bounds}

In this section we review various experimental and observational constraints on massless hidden photons \cite{Dobrescu:2004wz,Fabbrichesi:2020wbt} and discuss their relevance for our superhidden photons. The bounds are summarized in table \ref{tab:bounds} at the end of the section.

\subsection{Astrophysics and cosmology}
\label{sec:astro}

\paragraph{Star cooling: } Bremsstrahlung of superhidden photons from electrons in white dwarfs and red giants provides the most stringent astrophysical bound on the dipole coupling of the electron. These bounds originate from observational constraints on the maximum amount of energy loss allowed during star cooling.

The constraint is on a combination of the dipole coefficients\footnote{When $i$ and $j$ refer to the same flavor, we write only one index. For instance we write $d^{\rm M}_{e}\equiv d^{\rm M}_{ee}$ for the electron-electron coefficient, or $d^{\rm M}_{q}\equiv d^{\rm M}_{qq}$ for any quark-quark coefficient. Note also that, as can be seen from the definitions \eqref{eq:ReCImC}, $d_{ij}^{\rm M}$ and $d_{ij}^{\rm E}$ are complex quantities.} $d^{\rm M}_{ij}$, $d^{\rm E}_{ij}$ and the scale $\Lambda$. Translating the results of~\cite{Dobrescu:2004wz,Fabbrichesi:2020wbt} into this notation we have\footnote{We note that the more recent results of~\cite{Fabbrichesi:2020wbt} are given for the magnetic dipole coefficient only. Nevertheless, we suspect that this, as well as the other astrophysical and cosmological bounds, actually apply to the quantity $(|d^{\rm M}_{fg}|^2+|d^{\rm E}_{fg}|^2)^{1/4}$ (this is in line with~\cite{Dobrescu:2004wz}). A calculation of the limits including a full evaluation of the matrix elements of the processes (cf.~e.g.~\cite{Carenza:2021osu,Carenza:2025jwn}) would clarify this. To be conservative, in the main text we give the limits on the magnetic component. The same applies to the supernova limits discussed immediately below.
} 
\begin{equation}
\frac{\Lambda}{\sqrt{|d^{\rm M}_{e}}|}\gtrsim 1.3\times 10^3\text{ TeV}\,.
\end{equation}

\paragraph{Supernovae: } The quark-quark dipole coefficient is constrained from the measurement of the neutrino signal of the supernova 1987A~\cite{Dobrescu:2004wz,Fabbrichesi:2020wbt}. Recent limits from this are,
\begin{eqnarray}
\frac{\Lambda}{\sqrt{|d^{\rm M}_{q}|}}&\gtrsim & 4.1\times 10^2 \,\text{ TeV}\quad\,\text{~\cite{Fabbrichesi:2020wbt}}
\\\nonumber
&\gtrsim& 1\times 10^{3}\,\text{ TeV}\qquad \text{~\cite{Camalich:2020wac,Carenza:2019pxu}}\footnotemark \,.
\end{eqnarray}

\footnotetext{See also~\cite{Fiorillo:2025gnd} for a discussion of light particle production processes involving nucleons.}

Supernovae can also be used to constrain flavor non-diagonal couplings. For example, ref.~\cite{Camalich:2020wac} obtained a limit corresponding to
\begin{equation}
\frac{\Lambda}{\sqrt{|d_{ds}^{\rm M}|}}\gtrsim 1.2\times 10^3 \text{ TeV}\,.
\end{equation}

\paragraph{Big Bang Nucleosynthesis (BBN): } The coupling strength of the massless hidden photons has been argued to be constrained from primordial nucleosynthesis predictions and measurements \cite{Dobrescu:2004wz,Fabbrichesi:2020wbt}. Before recalling how this limit arises, let us note that in specific, realistic string realizations the decoupling temperature may lie above the temperature that can be treated safely in our model setup. Therefore, these limits do not directly apply in all cases. We will comment more on this in sect.~\ref{sec:bbncosmo}.

Keeping this caveat in mind, the usual story goes as follows~\cite{Dobrescu:2004wz,Fabbrichesi:2020wbt}: The energy density of relativistic species $\rho_{\rm R}$ at the BBN temperature $T_{\rm BBN}\approx 1$ MeV reads
\begin{equation}
    \rho_{\rm R}(T_{\rm BBN})=\frac{\pi^2}{30}\left[2+\frac{7}{2}+\frac{7}{4}\left(N_\nu+\Delta N_\nu\right)\right]T_{\rm BBN}^4\,,
    \label{dndef}
\end{equation}
where the factor $2$ counts the SM photon polarizations, the factor $\frac{7}{2}$ stands for the contribution of the electrons, $N_\nu$ counts the number of neutrino species and $\Delta N_\nu$ accounts for potential additional degrees of freedom, normalized like neutrinos. Experimental constraints from BBN place an upper bound on these additional degrees of freedom: $|\Delta N_\nu|\leq 0.278$ \cite{Fields:2019pfx}. Two polarizations from a hidden photon are too many and the bound would be violated if the photon was in thermal equilibrium at the BBN temperature. It should thus have decoupled earlier at a temperature $T_{\rm d}>T_{\rm BBN}$, and \emph{sufficiently} earlier for the contribution to the density at $T_{\rm BBN}$ to be compatible with the measurement errors.

To evaluate the hidden photon energy density at $T_{\rm BBN}$, we make use of entropy conservation:
\begin{equation}
  a^3T^3g_*(T)
  =\text{const.}
\end{equation}
Here $a$ is the scale factor and $g_*$ the number of relativistic degrees of freedom.\footnote{At our level of precision, it is sufficient to use $g_*$ instead of the more appropriate quantity $g_{*\text{S}}$ which accounts for the potentially different temperatures of different species.} This implies
\begin{equation}
    \rho_X(T_{\rm BBN})=\rho_X(T_{\rm d})\left(\frac{a(T_{\rm d})}{a(T_{\rm BBN})}\right)^4=\frac{2\pi^2}{30}\left[\frac{g_*(T_{\rm BBN})}{g_*(T_{\rm d})}\right]^\frac{4}{3}T_{\rm BBN}^4
\end{equation}
and, using \eqref{dndef}, one immediately obtains
\begin{equation}
     \Delta N_\nu(T_{\rm BBN})=\frac{8}{7}\left[\frac{g_*(T_{\rm BBN})}{g_*(T_{\rm d})}\right]^\frac{4}{3}\,.
\end{equation}
The upper bound on $\Delta N_\nu$ translates into a lower bound on $g_*(T_{\rm d})$. Assuming just the SM field content, this number can be translated into a temperature. The lowest allowed decoupling temperature, associated to the highest possible value for $\Delta N_\nu$, is then roughly $T_{\rm QCD}\approx 150$ MeV \cite{Dobrescu:2004wz}.

This lower bound on the decoupling temperature can further be translated into a constraint on the strength of the dipole interaction of the hidden photon. Indeed, at decoupling the interaction rate of the photon is of order of the Hubble scale, which allows one to express $T_{\rm d}$ in terms of the cross section of the hidden photon interactions with the SM. This cross section is directly related to the scale $\Lambda$ in the dipole interaction and the coupling strength $d_{ij}^{\rm M}$. One ends up with the lower bounds~\cite{Fabbrichesi:2020wbt}
\begin{equation}
\frac{\Lambda}{\sqrt{|d^{\rm M}_{e,\mu}|}}\gtrsim 51\text{ TeV}\,,\qquad\frac{\Lambda}{\sqrt{|d^{\rm M}_{q}|}}\gtrsim 41\text{ TeV}\,.
\end{equation}

\subsection{Dipole moments}

As anticipated earlier, the operator \eqref{eq:dipole} induces standard magnetic and electric dipole moments through the term involving $F'_{\mu\nu}$, suppressed by the kinetic mixing parameter $\epsilon$. Note that, in a setup where light SUSY partners are present, loop-suppressed diagrams involving them can also induce standard dipole moments. Depending on the values of the kinetic mixing parameter and the SUSY breaking scale as well as the characteristics of specific models, such contributions may compete with the tree-level effects induced by the hidden dipole operator. In what follows we assume that the latter dominate to derive our bounds, but this is to be taken with a grain of salt.

\subsubsection{Magnetic}
From \eqref{eq:dipole}, it follows that the magnetic dipole moment $D^{\text{M}}_{f}$ for fermions $f$ is
\begin{equation}
    D^{\text{M}}_{f}\equiv d^{\text{M}}_{f}\frac{v_{\rm h}\epsilon}{\Lambda^2}\,.
\end{equation}
This may be rewritten in terms of the quantity $\Delta a_{f}$ which is commonly used to characterize the anomalous magnetic moment:
\begin{eqnarray}
    \Delta a_{f}=\frac{\Delta(g_{f}-2)}{2}&=&d^{\text{M}}_{f}\frac{v_{\rm h}\epsilon m_{f}}{\Lambda^2}
    \\\nonumber
    &\sim& 1.3\times 10^{-14}\left(\frac{d^{\text{M}}_{f}}{1}\right)\left(\frac{\epsilon}{10^{-3}}\right)\left(\frac{100\,{\rm TeV}}{\Lambda}\right)^2\left(\frac{m_{f}}{m_{e}}\right)
    \\\nonumber
     &\sim& 2.6\times 10^{-12}\left(\frac{d^{\text{M}}_{f}}{1}\right)\left(\frac{\epsilon}{10^{-3}}\right)\left(\frac{100\,{\rm TeV}}{\Lambda}\right)^2\left(\frac{m_{f}}{m_{\mu}}\right)\,.
\end{eqnarray}
The experimental value of the electron magnetic moment has an uncertainty~\cite{Mohr:2024kco}
\begin{equation}
    \delta a_{e}(\text{exp})\sim 1.3\times 10^{-13}\,,
\end{equation}
implying the bound,
\begin{equation}
\frac{\Lambda}{\sqrt{|d^{\rm M}_{e}|}}\gtrsim 32 \text{ TeV}\left(\frac{\epsilon}{10^{-3}}\right)^{1/2}\,.
\end{equation}

For the muon magnetic moment, recent lattice computations suggest a good agreement between the SM prediction and the measured value \cite{Aliberti:2025beg},
\begin{equation}
    a_{\mu}(\text{exp})-a_{\mu}(\text{th})=(38\pm 63)\times 10^{-11}\,.
\end{equation}
Using just the uncertainty in this result yields a bound
\begin{equation}
\frac{\Lambda}{\sqrt{|d^{\rm M}_{\mu}}|}\gtrsim 6.4 \text{ TeV}\left(\frac{\epsilon}{10^{-3}}\right)^{1/2}\,.
\end{equation}
Note that, the previously reported discrepancy between experiment and theory \cite{Mohr:2024kco},
\begin{equation}
    a_{\mu}(\text{exp})-a_{\mu}(\text{th})=(253\pm 60)\times 10^{-11}\,,
\end{equation}
could be fit by a value $\Lambda/\sqrt{|d_\mu^{\rm M}|} \sim 3.2\text{ TeV}$.

\subsubsection{Electric}

In the presence of a CP violating term in the hidden-photon dipole operator, i.e.~for non-zero $d^{\text{E}}_{f}$, kinetic mixing also induces standard electric dipole moments (EDMs),
\begin{equation}
    D^{\text{E}}_{f}\equiv d^{\text{E}}_{f}\frac{v_{h}}{\Lambda^2}\epsilon \sim 
    1.7\times 10^{-24} e{\rm cm}\left(\frac{d^{\text{E}}_{f}}{1}\right)\left(\frac{\epsilon}{10^{-3}}\right)\left(\frac{100\,{\rm TeV}}{\Lambda}\right)^2\,.
\end{equation}
The currently best limit on the electron EDM is~\cite{Roussy:2022cmp},
\begin{equation}
|D^{\text{E}}_{e}|\leq 4.1\times 10^{-30}\,e\,{\rm cm}\,,
\end{equation}
which gives
\begin{equation}
\label{eq:bound_EDM}
\frac{\Lambda}{\sqrt{|d^{\rm E}_{e}|}}\gtrsim 6.4\times10^4 \text{ TeV}\left(\frac{\epsilon}{10^{-3}}\right)^{1/2}\,.
\end{equation}

\subsection{Flavor changing decays}

Considering flavor changing dipole moments between two fermion species $f$ and $g$,
\begin{equation}
\label{eq:dipoleflav}
    {\mathcal{L}}_{\rm dipole}=-\frac{v_{\rm h}}{2\Lambda^2}X_{\mu\nu} \left[ d^{\rm M}_{fg}\bar{f}\sigma^{\mu\nu}g+id^{\rm E}_{fg}\bar{f}\sigma^{\mu\nu}\gamma^{5}g\right] + \text{h.c.}
\end{equation}
and using the formulae from~\cite{Dobrescu:2004wz} (see also, e.g.~\cite{Barducci:2023hzo}), one has a decay rate
\begin{equation}
    \Gamma_{f\to g+X}=\frac{|d^{\rm M}_{fg}|^2+|d^{\rm E}_{fg}|^2}{16\pi}\frac{v^2_{\rm h}m^{3}_{f}}{\Lambda^4}\,.
\end{equation}
Specifically for muons this implies
\begin{equation}
    \Gamma_{\mu\to e+X}\sim 1.4\times 10^{-20}\,{\rm GeV}\frac{|d^{\rm M}_{\mu e}|^2+|d^{\rm E}_{\mu e}|^2}{1}\left(\frac{100\,{\rm TeV}}{\Lambda}\right)^4\,,
\end{equation}
which, together with the total muon decay rate, implies a branching ratio
\begin{equation}
    \text{BR}_{\mu\to e + X}\sim 0.05\frac{|d^{\rm M}_{\mu e}|^2+|d^{\rm E}_{\mu e}|^2}{1}\left(\frac{100\,{\rm TeV}}{\Lambda}\right)^4\,.
\end{equation}
This can be compared to the limit from~\cite{Jodidio:1986mz}
\begin{equation}
    \text{BR}_{\mu\to e + X}\leq 2.6\times 10^{-6}\,.
\end{equation}
We then infer the bound
\begin{equation}
\frac{\Lambda}{(|d^{\rm M}_{\mu e}|^2+|d^{\rm E}_{\mu e}|^2)^{1/4}}\gtrsim 1.2\times10^3 \text{ TeV}\,.
\end{equation}
A suitable search~\cite{Perrevoort} for this decay at the Mu3e experiment~\cite{Mu3e:2020gyw} may improve this limit by a factor $\sim 4$~\cite{Perrevoort}.

In presence of kinetic mixing we can also consider the induced decay $\mu\to e+\gamma$.
The corresponding rate is
\begin{equation}
    \text{BR}_{\mu\to e +\gamma}\sim 4.8\times 10^{-8}\frac{|d^{\rm M}_{\mu e}|^2+|d^{\rm E}_{\mu e}|^2}{1}\left(\frac{\epsilon}{10^{-3}}\right)^2\left(\frac{100\,{\rm TeV}}{\Lambda}\right)^4\,.
\end{equation}

Using the result of the recent MEGII experiment~\cite{MEGII:2025gzr} (for a previous similar bound from MEG see \cite{MEG:2016leq}),
\begin{equation}
\text{BR}_{\mu\to e + \gamma}\leq 1.5\times 10^{-13}\,,
\end{equation}
we obtain a limit
\begin{equation}
\label{eq:bound_flavor}
\frac{\Lambda}{(|d^{\rm M}_{\mu e}|^2+|d^{\rm E}_{\mu e}|^2)^{1/4}}\gtrsim 2.4\times 10^{3}\text{ TeV}\left(\frac{\epsilon}{10^{-3}}\right)^{1/2}\,.
\end{equation}
The final sensitivity for MEGII is expected to be $6\times 10^{-14}$ corresponding to a bound of $3.1\times 10^3\,{\rm TeV}$~\cite{MEGII:2025gzr} under the same conditions. Note, however, that similarly to the dipole moments, SUSY loop processes may contaminate this observable.

\subsection{Spin-dependent long-range forces}

The interaction \eqref{eq:dipoleint} also gives rise to spin-spin potentials~\cite{Dobrescu:2006au,Cong:2024qly}. Comparing with the forces from pseudoscalar-pseudoscalar exchange~\cite{Cong:2024qly},
parametrized in the literature by coupling constants 
$g^{f}_{\rm p}$, we obtain the identification\footnote{In~\cite{Cong:2024qly} the velocity-independent tensor-tensor potential differs from the pseudoscalar-pseudoscalar and the pseudotensor-pseudotensor potentials by a $\delta$ function. This should not affect the result for long-range force measurements.}
\begin{equation}
    -\frac{g^{f}_{\rm p}g^{f'}_{\rm p}}{4}\,\,\rightarrow\,\, 
\left(d^{\text{M}}_{f}d^{\text{M}}_{f'}+d^{\text{E}}_{f}d^{\text{E}}_{f'}\right)\frac{v^{2}_{\rm h}m_{f}m_{f'}}{\Lambda^4}\,.
\end{equation}
This turns into the formula
\bea
    \frac{\Lambda}{
\left[d^{\text{M}}_{f}d^{\text{M}}_{f'}+d^{\text{E}}_{f}d^{\text{E}}_{f'}\right]^{1/4}}\,\,\,
    \gtrsim\,\,\, (m_{f}m_{f'})^{1/4}\,\sqrt{2v_{\rm h}}\,(g^{f}_{\rm p}g^{f'}_{\rm p})^{-1/4}\,.
\eea
Let us concretely look at experiments testing forces between electron spins. In this case we have
\begin{eqnarray}
    \frac{\Lambda}{\left[(d^{\text{M}}_{e})^2+(d^{\text{E}}_{e})^2\right]^{1/4}} &\gtrsim& 3\times 10^{-2}\,{\rm TeV}\left(\frac{10^{-7}}{(g^{e}_{\rm p})^2}\right)^{1/4}
    \\\nonumber
        &\gtrsim&  5\,{\rm TeV}\left(\frac{10^{-16}}{(g^{e}_{\rm p})^2}\right)^{1/4}\,.
\end{eqnarray}
The value $(g^{e}_{\rm p})^2\simeq 10^{-7}$ used in the first line correspond to the sensitivity obtained in atomic measurements as done in~\cite{Ficek:2016qwp}.\footnote{In this case the measurement is between P states and the discrepancy in the $\delta$-function contribution should not contribute.} In the second line we use the value from a (macroscopic) spin source torsion pendulum~\cite{Heckel:2013ina,Terrano:2015sna}.

Spin-dependent forces between electrons and neutrons can also be constrained, cf. e.g.~\cite{Almasi:2018cob}, providing access to a different combination of the dipole interactions:
\begin{equation}
   \frac{\Lambda}{\left(d^{\text{M}}_{e}d^{\text{M}}_{n}+d^{\text{E}}_{e}d^{\text{E}}_{n}\right)^{1/4}}    \gtrsim 9\,{\rm TeV}\left(\frac{2\times10^{-14}}{g^{e}_{\rm p}g^{n}_{\rm p}}\right)^{1/4}\,.
\end{equation}

\begin{table}
\begin{center}
\scalebox{0.94}{
\begin{tabular}{c|c|c|c|c|c|c}
\backslashbox{Coeff.}{Exp.} & Stars & SN & BBN & Spin-dep. & \makecell{Flavor $\mu\to$\\ e+X, e+$\gamma$} & Dipole mom.\\\hline\hline
$\displaystyle{\frac{\Lambda}{\sqrt{|d_e^{\rm M}|}}}$ & $1.3\times 10^3$ &  & 51 &  & & $32\epsilon^{1/2}_{3}$\\\hline
$\frac{\Lambda}{\sqrt{|d_{\mu}^{\rm M}|}}$ &  & & 51 & & & $6.4\epsilon^{1/2}_{3}$\\\hline
$\displaystyle\frac{\Lambda}{\sqrt{|d_q^{\rm M}|}}$ & & $1\times 10^3$ & 41  & & & \\\hline
$\displaystyle\frac{\Lambda}{\sqrt{|d_{ds}^{\rm M}|}}$ & & $1.2\times 10^3$ &   & & & \\\hline
$\displaystyle\frac{\Lambda}{\sqrt{|d_e^{\rm E}|}}$ & & & &  & & $6.4\times 10^4\epsilon^{1/2}_{3}$\\\hline
$\displaystyle\frac{\Lambda}{\left(|d^{\rm M}_{\mu e}|^2+|d^{\rm E}_{\mu e}|^2\right)^{1/4}}$ &  & & & & \makecell{$1.2\times 10^3$\\ $2.4\times 10^3\epsilon^{1/2}_{3}$} & \\\hline
$\displaystyle\frac{\Lambda}{\left[(d^{\rm M}_{e})^2+(d^{\text{E}}_{e})^2\right]^{1/4}}$ &  & & & 5 & & \\\hline
$\displaystyle\frac{\Lambda}{\left(d^{\rm M}_{e}d^{\rm M}_{n}+d^{\text{E}}_{e}d^{\text{E}}_{n}\right)^{1/4}}$ &  & & & 9 & & \\
\end{tabular}
}
\end{center}
\caption{Bounds on the dipole couplings (in TeV) from various processes. The parameter $\epsilon_{3}\equiv\epsilon/10^{-3}$ indicates the scaling with $\epsilon$, normalized to a typical value $10^{-3}$.}
\label{tab:bounds}
\end{table}

\section[Implications of phenomenology for string theory and vice versa]{\texorpdfstring{Implications of phenomenology for string theory and\\[.1cm]
vice versa}}
\label{sec:pheno}

\subsection{Superhidden photons in plain sight?}
\label{sec:tableMS}

In the context of string compactification with a large volume, the absence of deviations from 4d Newtonian gravity implies\footnote{$\Mp\simeq 2.4\times 10^{18}$~GeV is the reduced Planck mass.} $m_{\rm KK}\gtrsim 10^{-30}\Mp$ (cf.~\cite{Kapner:2006si,ParticleDataGroup:2018ovx}), which is a very weak constraint. Indeed, this allows for volumes as large as $10^{45}$ and a string scale as low as $10^{-22.5}\Mp$ in the isotropic case. For such models the LHC provides a more stringent constraint, excluding string scales $\sim$~TeV or below (cf.~e.g.~\cite{ParticleDataGroup:2018ovx,ATLAS:2014gys,Landsberg:2015pka}). From this perspective, the bounds on superhidden photon couplings collected in table~\ref{tab:bounds} exclude several orders of magnitude of a priori allowed parameter space, as we will see shortly.  In anisotropic setups with two large extra dimensions \cite{Antoniadis:1998ig}, the LHC bound competes with tests of 4d gravity while in scenarios with a single extra dimension \cite{Montero:2022prj}, submillimeter gravity becomes the strongest constraint. In this specific case, the gravity constraint implies $\Ms\gtrsim 10^5$ TeV, but our bounds inferred from the dipole operator can be even more stringent as we will see momentarily.

We now formulate a main result of this work, namely the translation of the experimental constraints summarized in table~\ref{tab:bounds} into lower bounds on the string scale. For this purpose, we employ the expression for the scale $\Lambda$ suppressing the dipole operator in the brane photons case, cf.~eq.~\eqref{eq:exp_Lambda_brane}. In addition, we take into account the three flavor-structure scenarios introduced in sect.~\ref{sec:yuk}. As a result, one obtains lower bounds on $\Ms$ as presented in table~\ref{tab:boundsMs} for kinetic-mixing-independent and in table~\ref{tab:boundsMs_kinmix} for kinetic-mixing-dependent observables. As an example, let us consider the most stringent bound in table~\ref{tab:bounds}, namely that from the electric dipole moment of the electron: $\Lambda/\sqrt{|d_e^{\rm E}|}\gtrsim 6.4\times 10^4\epsilon_3^{1/2}$~TeV. Upon using \eqref{eq:exp_Lambda_brane}, which relates $\Lambda$ to $\Ms$, one gets $\Ms/\sqrt{|d_e^{\rm E}|}\gtrsim 3\times 10^6\epsilon_3^{1/2}$ TeV, assuming a D3-brane located one string length away from the SM. Without any flavor suppression, i.e. $d_e^{\rm E}\sim 1$, and for a loop-suppressed kinetic mixing parameter, i.e. $\epsilon_3\sim 10^{-3}$, the obtained numerical value is the one reported in the top left corner of table~\ref{tab:boundsMs_kinmix}. The weaker values on the same line are then simply obtained by applying a flavor suppression $d_e^{\rm E}\sim\varepsilon\approx 5.9\times 10^{-2}$ or $d_e^{\rm E}\sim\varepsilon'\approx 2.9\times 10^{-4}$. All other entries in tables~\ref{tab:boundsMs} and \ref{tab:boundsMs_kinmix} are obtained in exactly the same manner\footnote{Note that for the spin-dependent forces we use the strongest constraint of table~\ref{tab:bounds} coming from electron-neutron interactions. The neutron coefficients $d_{n}^{\rm M,E}$ can be approximately identified with $d_u^{\rm M,E}$ or $d_d^{\rm M,E}$ \cite{Dobrescu:2004wz}, which we further approximate by the leptonic coefficients $d_e^{\rm M,E}$ for simplicity. We also assume no interference in this case. Note also that, when using the results of table~\ref{tab:bounds} and assuming that $d^{\rm M,E}\sim{\cal O}(1)$, this introduces a factor $2^{1/4}$ for some of the bounds, e.g.~$((d^{\rm M}_{e})^2+(d^{\text{E}}_{e})^2)^{1/4}\,\to\,2^{1/4}$. This effect is, however, irrelevant for our level of precision. Indeed, the two-digit accuracy of the quoted values should be interpreted with care.}.

\begin{table}
\begin{center}
\caption*{Kinetic-mixing-independent constraints from brane photons.}
\begin{tabular}{c|c|c|c}
Exp. & $d_{\mu e}^{\rm M,E}\sim d_e^{\rm M,E}\approx\mathcal{O}(1)$ &  $d_{\mu e}^{\rm M,E}\sim d_e^{\rm M,E}\sim\varepsilon\approx 0.059$ & $d_e^{\rm M, E}\sim\varepsilon'\approx 0.00029$\\\hline\hline
\makecell{FCNC\\ $\mu\to e+X$} & $\Ms\gtrsim 6.8\times 10^4$ TeV & $\Ms\gtrsim 1.6\times 10^4$ TeV & \\\hline
Stars & $\Ms\gtrsim 6.2\times 10^4$ TeV & $\Ms\gtrsim 1.5\times 10^{4}$ TeV & $\Ms\gtrsim 10^3$ TeV\\\hline
Spin-dep. pot. & $\Ms\gtrsim 5.1\times 10^2$ TeV & $\Ms\gtrsim 1.2\times 10^2$ TeV & $\Ms\gtrsim 9$ TeV\\
\end{tabular}
\end{center}
\caption{Lower bounds on the string scale depending on the flavor structure of the dipole operator for the three observables independent on kinetic mixing ($\varepsilon$ and $\varepsilon'$ denote the flavor structure suppression as specified in eqs.~\eqref{eq:flavour} and \eqref{eq:flavour2}): Flavor Changing Neutral Currents (FCNC) through the process $\mu\to e + X$, star cooling and the spin-dependent potentials. The corresponding RR photon bounds are weaker by a factor $\simeq 100$.}
\label{tab:boundsMs}
\end{table}

\begin{table}
\begin{center}
\caption*{Kinetic-mixing-dependent constraints from brane photons.}
\begin{tabular}{c|c|c|c}
Exp. & $d_{\mu e}^{\rm M,E}\sim d_e^{\rm M,E}\approx\mathcal{O}(1)$ &  $d_{\mu e}^{\rm M,E}\sim d_e^{\rm M,E}\sim\varepsilon\approx 0.059$ & $d_e^{\rm M, E}\sim\varepsilon'\approx 0.00029$ \\\hline
\multicolumn{4}{c}{$\epsilon\sim 10^{-3}$}\\\hline\hline
$e^-$ EDM & $\Ms\gtrsim 3\times 10^6$ TeV & $\Ms\gtrsim 7.4\times 10^5$ TeV & $\Ms\gtrsim 5.2\times 10^4$ TeV\\\hline
\makecell{FCNC\\ $\mu\to e+\gamma$} & $\Ms\gtrsim 1.4\times 10^5$ TeV & $\Ms\gtrsim 3.3\times 10^4$ TeV & \\\hline
\multicolumn{4}{c}{$\epsilon\sim \mathcal{O}(1)$}\\\hline\hline
$e^-$ EDM & $\Ms\gtrsim 9.5\times 10^7$ TeV & $\Ms\gtrsim 2.3\times 10^7$ TeV & $\Ms\gtrsim 1.6\times 10^6$ TeV\\\hline
\makecell{FCNC\\ $\mu\to e+\gamma$} & $\Ms\gtrsim 4.3\times 10^6$ TeV & $\Ms\gtrsim 10^6$ TeV & \\
\end{tabular}
\end{center}
\caption{Lower bounds on the string scale depending on the flavor structure of the dipole operator for the two kinetic-mixing-dependent experimental observables: The electric dipole moment of the electron and FCNC through the process $\mu\to e+\gamma$. The upper part of the table assumes a loop-suppressed kinetic mixing $\epsilon\sim 10^{-3}$ while the lower part assumes an order one mixing. The corresponding RR photon bounds are weaker by a factor $\simeq 100$.}
\label{tab:boundsMs_kinmix}
\end{table}

For RR photons, the relevant relation between $\Lambda$ and the string scale is given in eq.~\eqref{eq:exp_Lambda_RR}. By comparing with the brane photon relation, we see that the scale $\Lambda$ is enhanced by a factor of $2^5\pi\approx 100$. Consequently, the bounds on the string scale can be read directly from the same tables \ref{tab:boundsMs} and \ref{tab:boundsMs_kinmix} by dividing the entries by this factor, which leads to correspondingly weaker constraints.

As noted earlier, concerning the electron EDM and the kinetic-mixing-dependent flavor-changing decay $\mu\to e+\gamma$, competing SUSY loop contributions may contaminate these observables. However, the process $\mu\to e+X$ and the astrophysical bounds from stellar cooling do not depend on the kinetic mixing and they both provide constraints of comparable magnitude as the $\mu\to e+\gamma$ bound. Spin-dependent force experiments yield weaker limits but remain relevant in light of the potential for significant future improvements in sensitivity.

The bounds discussed so far can already exclude a non-trivial range of values for the string scale in models with large compactification volumes. These limits can, however, be pushed even further. Indeed, a loop-suppressed kinetic mixing is not the generic expectation for branes wrapping relatively small cycles, where couplings are typically of order unity. If the kinetic mixing parameter $\epsilon$ is taken to be order one, i.e. $\epsilon_3\sim 10^3$, the electron EDM constraint \eqref{eq:bound_EDM} and the bound from $\mu\to e+\gamma$ in \eqref{eq:bound_flavor} are significantly strengthened,
\begin{equation}
\frac{\Lambda}{\sqrt{|d_e^{\rm E}}|}\Bigg\rvert_{\epsilon=1}\gtrsim2\times 10^6 \text{ TeV}\,,\qquad\frac{\Lambda}{\left(|d^{\rm M}_{\mu e}|^2+|d^{\rm E}_{\mu e}|^2\right)^{1/4}}\Bigg\rvert_{\epsilon=1}\gtrsim 7.6\times 10^4 \text{ TeV}\,.
\end{equation}
These estimates lead to the lower part of table~\ref{tab:boundsMs_kinmix}, following the same procedure described above. The bounds from RR photons can again be obtained by dividing the table entries by a factor $2^{5}\pi\approx 100$.

Finally, as mentioned earlier, if particles lighter than the KK modes but with the same interactions are present in a model, they would induce a dipole coupling with a much less suppressed energy scale and yield stronger constraints on the string scale. It would be interesting to investigate such scenarios more carefully in the future.

\subsection{Hide and seek in LVS}
\label{sec:LVS}

We now refine the results of the previous subsection by considering bounds on the string scale arising in a specific moduli stabilization scenario, the LVS \cite{Balasubramanian:2005zx,Conlon:2005ki}. It turns out that the \emph{F-term problem} \cite{Hebecker:2019csg} limits the extent to which  the suppression scale of the dipole operator can be lowered. We discuss how this compares with experimental constraints. 

\subsubsection{LVS in a nutshell}

In type IIB string theory compactifications on CY orientifolds with fluxes and O3/O7-planes \cite{Giddings:2001yu, Grana:2005jc, Denef:2008wq},
the tree-level scalar potential for Kahler moduli is no-scale. The interplay of non-perturbative, $\alpha'$ and loop corrections allows for stabilization of the Kähler moduli at an exponentially large overall volume $\V$ \cite{Balasubramanian:2005zx,Conlon:2005ki}.

To be more specific, after stabilization of the complex-structure moduli and the axio-dilaton by fluxes at high scale \cite{Giddings:2001yu}, one is left with a constant superpotential $W=W_0$ and a Kähler poential $K=-2\ln\V$. The no-scale structure arises because the volume $\V$ is a homogeneous function of degree $3/2$ in the complexified Kähler moduli $T^i$, $i=1,\dots,h^{(1,1)}_+$, such that
\begin{equation}
    \kappa_4^2V_{\rm tree}=e^K\left(K^{i\bar \jmath}D_iW_0 D_{ \bar \jmath}\overbar{W}_{\!0}-3|W_0|^2\right)=0\,.
\end{equation}
Here $D_iW_0\equiv(\partial_i+K_i)W_0=K_i W_0$, $K_i\equiv\partial_iK$ and 
\begin{equation}
\kappa_4\equiv\frac{1}{\Mp^2}\,. 
\end{equation}
The two leading effects relevant for Kähler moduli stabilization are:
\begin{itemize}
    \item Non-perturbative corrections from gaugino condensation on D7-branes or Euclidean D3-brane instantons which modify the superpotential to
    \begin{equation}
        W=W_0+A_ie^{-a_iT^i}\,.
    \end{equation}
    \item $\alpha'$ corrections from the 10d action which induce a correction to the Kähler moduli Kähler potential \cite{Becker:2002nn},
    \begin{equation}
        K=-2\ln\left(\V+\frac{\xi}{2\gs^{3/2}}\right)\qquad\mbox{with}\qquad\xi\equiv(h^{(2,1)}-h^{(1,1)})\frac{\zeta(3)}{2(2\pi)^3}\,.
        \end{equation}
\end{itemize}

The simplest implementation of this scenario involves a swiss-cheese geometry with one large four-cycle $\tau_{\rm b}$ and a small four-cycle $\tau_{\rm s}$ such that the volume is given by $\V\approx\tau_{\rm b}^{3/2}-\tau_{\rm s}^{3/2}\sim\tau_{\rm b}^{3/2}$. Omitting order-one factors, the scenario yields
\begin{equation}
\tau_{\rm s}\sim\frac{1}{\gs}\,,\qquad\qquad\V\sim\gs^{-1/2
}e^{\frac{\mathcal{O}(1)}{\gs}}\,.
\end{equation}

\begin{figure}
    \centering
    \includegraphics[scale=0.5]{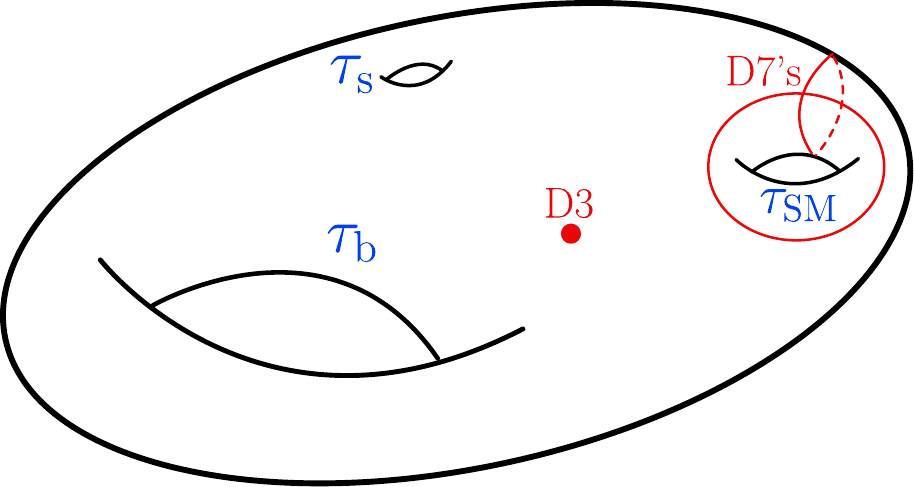}
    \caption{Geometric picture of our setup in the LVS scenario. The cycles $\tau_{\rm b}$ and $\tau_{\rm s}$ schematically represent respectively the large and small cycles in a swiss-cheese realization of the LVS. Stacks of D7-branes wrap relatively small cycles, collectively denoted $\tau_{\rm SM}$, and the SM is realized on their intersection. An isolated D3-brane sits somewhere in the internal space.}
    \label{fig:geom2}
\end{figure}

In this setup, as before, one can think of the SM as being realized on intersecting stacks of D7-branes wrapping additional (relatively small) four-cycles collectively denoted $\tau_{\rm SM}$. Finally, we may add a spacetime-filling D3 brane. This is illustrated in figure \ref{fig:geom2}, which updates figure \ref{fig:geom} for the case at hand.

\subsubsection{Low string scale in the LVS}\label{sec:LoLVS}

As we have seen, our dimension-six dipole operator is suppressed by $\Lambda^2$, where $\Lambda$ is set by the string scale $\Ms$ up to numerical factors (see \eqref{eq:exp_Lambda_brane}, \eqref{eq:exp_Lambda_RR}). To evaluate whether this operator can be accessible to experiments, it is crucial to understand how small the string scale can be in the LVS.

\paragraph{Scaling of potential and masses:}
We recall that the two contributions to the LVS potential, $V_{\rm LVS}\sim\delta V_{\rm np}+\delta V_{\alpha'}$, scale as~\cite{Balasubramanian:2005zx},
\begin{equation}
\kappa_4^2\,\delta V_{\rm np}\sim\frac{W_0^2}{\V^3}\log^{3/2}(W_0/\V)\qquad\mbox{and}\qquad \kappa_4^2\,\delta V_{\alpha'}\sim\frac{W_0^2}{\V^3}\,.
\end{equation}
Here, we omit numerical and $\gs$ factors since they are irrelevant at our level of precision. The masses of gravitino, volume modulus and the lightest KK modes read
\begin{equation}
m_{3/2}\sim\frac{W_0}{\V}\Mp\,,\qquad  m_\V\sim\frac{W_0}{\V^\frac{3}{2}}\Mp\,,\qquad m_{\rm KK}\sim\frac{\Mp}{\V^{\half+\frac{1}{l}}}\,,
\end{equation}
where $l$ is the number of the parametrically largest dimensions within the 6d internal space. In the isotropic case with $l=6$, one recovers the standard $\V^{-2/3}$ scaling of $m_{\rm KK}$.

\paragraph{Phenomenological requirements:}
To understand how low $\Ms$ can be, we have to keep in mind three phenomenological requirements:
\begin{enumerate}[1)]
\item $m_{1/2}\gtrsim 10^{-15}\Mp$, such that superpartners of SM fields are heavier than $\sim 1$~TeV.
\item $m_{\rm KK}\gtrsim 10^{-30}\Mp$, from bounds on deviation from 4d Newtonian gravity.
\item $m_\V\gtrsim 10^{-30}\Mp$, to avoid a fifth force mediated by the volume modulus which couples to matter with $1/\Mp$ suppression.
\end{enumerate}

\paragraph{Other moduli:}

The small cycle $\tau_{\rm s}$ is stabilized with a modulus mass $m_{\tau_{\rm s}}\sim(W_0/\V)\Mp$ which is large and thus irrelevant for us. Some of the other Kähler moduli may be stabilized by loop corrections, with masses smaller than that of the volume. This is, however, not a problem since their coupling to matter is suppressed as $m_{3/2}/\Mp^2$ and thus does not induce significant fifth forces~\cite{Cicoli:2012tz}.

\paragraph{Masses of superpartners:}

Depending whether the SM is realized on (fractional) D3- or D7-branes, the gaugino mass reads
\begin{equation}
\text{For D3: }\,\,\,m_{1/2}\sim\frac{W_0}{\V^2}\Mp\,\,,\qquad\qquad\qquad\text{For D7: }\,\,\,m_{1/2}\sim\frac{W_0}{\V}\Mp\,.
\end{equation}
The difference comes from the gauge kinetic function involved in the computation of the mass. In the first case it is given by the axio-dilaton $S$ and in the second case by a Kähler modulus $T$. The D7 case is then favored, but it is still problematic as we will now see.

We can rewrite the mass scales in terms of $\Ms=\Mp/\V^\half$ to get
\begin{equation}
m_{1/2}\sim m_{3/2}\sim W_0\left(\frac{\Ms}{\Mp}\right)^2\!\Mp\,,\quad m_\V\sim W_0\left(\frac{\Ms}{\Mp}\right)^3\!\Mp\,,\quad m_{\rm KK}\sim\left(\frac{\Ms}{\Mp}\right)^{1+\frac{2}{l}}\!\Mp\,.
\end{equation}
In this situation, requirement 1) implies
\begin{equation}
\label{eq:bound_LVS}
W_0\left(\frac{\Ms}{\Mp}\right)^2\gtrsim 10^{-15}\quad\Longleftrightarrow\quad\Ms\gtrsim \frac{10^{-7.5}}{\sqrt{W_0}}\Mp\sim\frac{10^{7.5}}{\sqrt{W_0}}\text{ TeV.} 
\end{equation}
Such a relatively high string scale is of the same order as the strongest bounds reported in sect.~\ref{sec:tableMS}, implying that we lack sensitivity with present data.

\paragraph{Stronger SUSY breaking:}

To lower the string scale further, one needs stronger SUSY breaking effects than those coming from Kähler moduli. Following \cite{Hebecker:2019csg}, we consider a toy model where some field $X$ with a positive $F$-term $F$ breaks supersymmetry. Higher-dimension operators suppressed by a scale $M$ then induce soft terms, in particular $m_{1/2}\sim F/M$. The corresponding contribution to the scalar potential is
\begin{equation}
\delta V_X\sim F^2\sim M^2 m_{1/2}^2\,.
\end{equation}

To avoid an uplift exceeding today's almost vanishing vacuum energy, $\delta V_X$ may not be larger than the depth of the LVS AdS minimum. This restriction has been discussed under the name ``$F$-term problem'' in \cite{Hebecker:2019csg}. It is quantified by
\begin{equation}
\delta V_X\lesssim\frac{W_0^2}{\V^3}\,\Mp^4 \quad\Longleftrightarrow\quad\frac{\Ms}{\Mp}\gtrsim\left(\frac{Mm_{1/2}}{\Mp^2W_0}\right)^\frac{1}{3}\,.
\end{equation}
Making the most optimistic assumption about the SUSY breaking sector, $M\sim m_{1/2}\sim\,$TeV, requirement 1) then implies
\begin{equation}
\frac{\Ms}{\Mp}\gtrsim\frac{10^{-10}}{W_0^\frac{1}{3}}\quad\Longleftrightarrow\quad\Ms\gtrsim  \frac{10^5}{W_0^\frac{1}{3}}\text{ TeV.}
\end{equation} 
Plugging this into the expressions for $m_\V$ and $m_{\rm KK}$ we obtain
\begin{equation}
m_\V\gtrsim 10^{-30}\Mp\,,\qquad m_{\rm KK}\gtrsim\frac{10^{-10-\frac{20}{l}}}{W_0^{\frac{1}{3}+\frac{2}{3l}}}\Mp\,,
\end{equation}
which is in agreement with requirement 2) and (marginally) with requirement 3). We observe that the lower bound on $\Ms$ is slightly reduced compared to \eqref{eq:bound_LVS}, such that present experimental sensitivities summarized in tables~\ref{tab:boundsMs} and \ref{tab:boundsMs_kinmix} can already provide non-trivial constraints.

\subsubsection{Cosmology}
\label{sec:bbncosmo}

As already mentioned in sect.~\ref{sec:astro}, we expect BBN constraints on superhidden photons not to affect realistic stringy vacua, because the hidden photons decouple at a much higher temperature than  can realistically be reached. Thus, their abundance hence becomes negligible. Indeed, we can estimate the cross section induced by the dipole interaction as
\begin{equation}
    \sigma\sim\frac{v_{\rm h}^2}{\Lambda^4}\,.
\end{equation}
Requiring the rate $\Gamma\sim T_{\rm d}^3\sigma$ to be of the order of $H(T_{\rm d})\sim T_d^2/\Mp$ and using $\Lambda\sim \Ms$ gives
\begin{equation}
    T_{\rm d}\sim\Mp\left(\frac{\Mp}{v_{\rm h}}\right)^2\frac{1}{\V^2}\,.
\end{equation}
This decoupling temperature should be compared with the temperature of moduli stabilization $T_{\rm mod}\sim m_{\rm mod}\sim\frac{\Mp}{\V^{3/2}}$. The volume scalings are such that $T_{\rm d}$ could in principle be below the moduli temperature but the huge factor $(\Mp/v_{\rm h})^2$ cannot realistically be overcome. In addition, at high energy a more relevant estimate of the cross section is
\begin{equation}
    \sigma\sim\frac{T^2}{\Lambda^4}\quad\Longrightarrow\quad T_{\rm d}\sim\frac{\Mp}{\V^\frac{2}{3}}\,.
\end{equation}
Based on this, it is not even possible in principle to push $T_{\rm d}$ below $T_{\rm mod}$.

\section{Towards a photinoverse}
\label{sec:photinoverse}

As mentioned in the introduction, once the field redefinition of \eqref{eq:rotation} is performed, a superhidden photon decouples from the SM at the renormalizable level. However, this is not necessarily true for its SUSY partner $\lambda_X$. Indeed, depending on the particular details of the SUSY breaking sector, the superhidden photinos mix with the MSSM neutralinos. This was initially noticed and analyzed in \cite{Ibarra:2008kn,Arvanitaki:2009hb,Goodsell:2011wn}. In this section we will quote these results and reinterpret them for our particular scenario. Note, however, that a direct translation is not possible: As will become apparent shortly, the mass ranges considered in \cite{Ibarra:2008kn,Arvanitaki:2009hb,Goodsell:2011wn} differ from our expectations for the superhidden photinos. Moreover, also the cosmological history might be modified compared to these references. Our results should therefore be taken as rather preliminary.

The relevant part of the action is (in two-component Weyl spinor notation) 
\be
e^{-1}\mathcal L=-\frac14\Re\mathcal K_{ab}F^{a,\mu\nu}F^{b}_{\mu\nu}-i\Re\mathcal K_{ab}\overline\lambda^{a}\overline\sigma^\mu\partial_\mu\lambda^{b}+\big(\mathcal M_{ab}\lambda^{a}\lambda^{b}+\text{h.c.}\big)+\cdots
\ee
Here, and in what follows $F^{a,\mu\nu}\equiv\big(F^{\mu\nu},X^{\mu\nu}\big)$ denotes the SM and superhidden photon field strengths and $\lambda^{a}\equiv\big(\lambda_{\rm SM},\lambda_X\big)$ stands for their respective gauginos. $\mathcal K$ is the matrix of gauge kinetic functions:
\be
\mathcal K=\begin{pmatrix}f_{\rm SM}&\epsilon\\\epsilon&f_X\end{pmatrix}.
\ee
If the SM is realized on D7-branes, the corresponding gauge kinetic function is simply the SM four-cycle K\"ahler modulus: $f_{\rm SM}=T_{\rm SM}/2\pi$. On the other hand, $f_X$ is either $f_X=S/2\pi$ in the case of a D3, or a more involved holomorphic function $f_X(U_i)$ of the complex structure moduli $U_i$ in the case of RR photons \cite{Grimm:2004uq}. As before, $\epsilon$ denotes the kinetic mixing parameter. It is natural to expect that $\epsilon$ depends on $U_i$ and $S$, but to the best of our knowledge the precise dependence is not known in the generic case.\footnote{For brane photons, the findings of \cite{Hebecker:2023qwl} suggest that $\epsilon$ may also depend on K\"ahler moduli. However, it is an open question how this is compatible with the holomorphic structure of gauge kinetic functions and the discrete shift symmetry of K\"ahler moduli \cite{Bullimore:2010aj}.} 

The gaugino mass matrix explicitly depends on SUSY breaking and its mediation. In the LVS we have
\be
\mathcal M=\frac{F^i}{\Mp}\partial_i\mathcal K\,,\qquad \kappa_4F^{i}=e^{K/2}K^{i\overline\jmath}D_{\overline \jmath}\overline W\,.
\ee
In general, diagonalizing $\mathcal K$, e.g. with a transformation similar to  \eqref{eq:rotation} accompanied by a rescaling, will not make  $\mathcal M$ diagonal as well. We denote the resulting mass matrix by $\hat{\mathcal M}$,
\be
\hat{\mathcal M}=\begin{pmatrix}\mathcal M_A&\delta\mathcal M\\\delta\mathcal M&\mathcal M _X\end{pmatrix}\simeq\frac1{\Mp}\begin{pmatrix}F^{T_{\rm SM}}&F^{S,U}\partial_{S,U} \epsilon-\epsilon F^{S,U}\partial_{S,U} f_X\\
F^{S,U}\partial_{S,U} \epsilon-\epsilon F^{S,U}\partial_{S,U} f_X&F^{S,U}\partial_{S,U}f_X\end{pmatrix}.\label{eq:Mhat}
\ee
In the above we have assumed small kinetic mixing and set gauge couplings to unity. After SUSY and electroweak symmetry breaking, the bino, the two neutral higgsinos and the wino mix. Since $\delta \mathcal M$ is generically non-zero, these MSSM neutralinos also mix with $\lambda_X$, inducing interactions between the SM and the superhidden photino. On the one hand this allows for direct searches at colliders but on the other hand, if $\mathcal M_X$ is smaller than the masses of the MSSM sparticles this can have significant influence on the evolution of the universe. A thorough estimation of $\hat{\mathcal M}$ and its phenomenological consequences, although highly relevant as potentially very constraining, is beyond the scope of this paper and left for future work. We will only provide a crude estimate of $\hat{\mathcal M}$ and draw preliminary conclusions in the light of results from \cite{Ibarra:2008kn,Arvanitaki:2009hb,Goodsell:2011wn}.

As can be seen from \eqref{eq:Mhat}, $\hat{\mathcal M}$ can be estimated, provided knowledge of $F^{T_{\rm SM}}$, $F^U$, and $F^S$. In the LVS one expects \cite{Conlon:2005ki,Blumenhagen:2009gk,Aparicio:2014wxa}
\be
F^{T_{\rm SM}}\sim\frac{\Mp^2}{\vol}\,,\qquad F^S\sim\frac{\Mp^2}{\vol^2}\qquad\text{ and }\qquad F^U\sim\frac{\Mp^2}{\vol^2}\,.
\ee
The above estimate for $F^{T_{\rm SM}}$ agrees with what one would expect at the level of GKP. By contrast, the two other $F$-terms would vanish at this level of precision since \mbox{$D_{S,U}W=0$}.
The values of $F^S$, $F^U$ given above come from subleading effects: From non-perturbative superpotential corrections $W_{\rm np}$ and 
from corrections to the K\"ahler potential. The latter enter the analysis is two ways: First, in the form of non-diagonal entries in the inverse K\"ahler metric (e.g. $K^{U\overline T}\neq0$) and, second, through the modification of the K\"ahler-covariant derivatives $D_{S,U}$. So far, only the $W_{\rm np}$ contribution to $F^U$ was computed in \cite{Aparicio:2014wxa}. We have explicitly checked (cf.~appendix \ref{app:Fterm}) that corrections to the K\"ahler potential lead to the same volume scaling of $F^U$. Note that additional contributions, either from a local source of SUSY breaking, as in section \ref{sec:LoLVS}, or from the uplift to dS are possible. For example, if some non-ISD flux is used for uplifting,\footnote{Such 
$F$-term uplifts~\cite{Saltman:2004sn}
have attracted growing attention over recent years (see e.g.~\cite{
Gallego:2017dvd,Honma:2019gzp,Hebecker:2020ejb, Krippendorf:2023idy,
Lanza:2024uis}), which may be well justified given the control problems of the anti-D3-brane. Of course, they very much rely on the tuning power of the flux Landscape, which one may also question~\cite{Bena:2020xrh}.
} 
one expects $F^{U,S}\sim \Mp^2 \vol^{-3/2}$ \cite{Conlon:2005ki}. Again, we postpone the discussion of such complications to future work.

A key quantity for phenomenology is the
mixing angle $\theta$ between a bino and a superhidden photino:
\be
\frac12\tan(2\theta)\equiv \frac{\delta \mathcal M}{\mathcal M_A-\mathcal M_X}\sim\frac{F^{S,U}\partial_{S,U} \epsilon-\epsilon F^{S,U}\partial_{S,U}f_X}{F^{T_{\rm SM}}}\,.
\ee
As one expects $\partial_{S,U}\epsilon\sim\mathcal O(\epsilon)$, it is convenient to define $f_{{S,U};\epsilon}\equiv(\partial_{S,U}\epsilon)/\epsilon\sim\mathcal O(1)$ to get
\be
\theta\sim \epsilon\frac{F^{S,U}(f_{{S,U};\epsilon}-\partial_{S,U}f_X)}{F^{T_{\rm SM}}}\sim\frac{\epsilon}\vol\,.\label{eq:theta_mix}
\ee
If $\theta$ is sufficiently large and the reheating temperature is high enough for MSSM neutralino production, superhidden photinos can be produced thermally and may overclose the universe. Whether this is the case or not is model dependent. For instance, in the case of K\"ahler moduli inflation with a final volume $\vol\sim10^{10}$, one can have reheating temperatures of order $T_{\rm rh}\sim1\,{\rm GeV}$ with the DM abundance being acounted for by an axion \cite{Cicoli:2022fzy}. In scenarios of this type the reheating temperature might not be high enough to produce the MSSM LSP, hence the following cosmological bounds are circumvented.

By contrast, in the case of sufficiently high $T_{\rm rh}$, the bound from thermal $\lambda_X$ production and overclosure is \cite{Ibarra:2008kn}
\be
\theta\lesssim5\times10^{-12}\sqrt{\frac{\mathcal M_A}{\mathcal M_X}}\,.\label{eq:overclose}
\ee
Using \eqref{eq:Mhat} and \eqref{eq:theta_mix}, this constrains the volume and the kinetic mixing:
\be
\epsilon^{2/3}\,3\times10^7\lesssim \vol\lesssim 10^{15}\,,\label{eq:photino_reheat}
\ee
where the upper bound comes from requiring the SUSY scale in the SM sector to be above $1\,\mathrm{TeV}$.

Within the range given in equation \eqref{eq:photino_reheat}, additional restrictions are expected: If R-parity is conserved, the lightest MSSM sparticle is stable up to decay into $\lambda_X$. As the volume $\vol$ grows, $\theta$ decreases and the decay rate $\Gamma_{{\rm MSSM}\rightarrow \lambda_X+{\rm SM}}$ from the MSSM to a hidden gaugino and a SM particle becomes more and more suppressed. For example, if the lightest MSSM sparticle is a slepton one expects 
\begin{equation}
\Gamma_{{\rm MSSM}\rightarrow \lambda_X+{\rm SM}}\sim\vol^{-3}\,.
\end{equation}
Provided the freeze-out abundance of the lightest MSSM sparticle is not too large, this is not necessarily problematic, especially at the lower end (small volumes) of \eqref{eq:photino_reheat}, as $\lambda_X$ can make up a part of dark matter. However, as one considers larger volumes two potential issues arise: First, since the hierarchy between hidden and observable gaugino masses is considerable, the hidden photinos inherit a large amount of kinetic energy which in turn leads to potentially long free streaming lengths $\lambda_{\rm FS}$. Those are constrained by structure formation. Second, for some range of volumes (typically at the upper end of the range \eqref{eq:photino_reheat}) the decays can happen during BBN, therefore potentially spoiling the successful prediction of abundances of light elements. Summarizing, we conclude that the photinoverse has the potential to severely constrain the CY volume and the reheat temperature but also provides potentially interesting new observable signatures.

\section{Conclusions}
\label{sec:conclusion}

The abundance and genericity of hidden photons in the string theory Landscape makes them excellent probes for new physics. Similar to the string theory \emph{axiverse}, the string theory \emph{photoverse} offers a rich playground in which to look for potential signals of string theory and its models. We argued that the hidden photons are often massless and do not couple to any light dark current such that they have no renormalizable direct couplings to SM fields: This defines what we called \emph{superhidden photons}.

The leading interaction of superhidden photons is through a dimensions-six operator that couples two SM fermion fields, the hidden photon and the Higgs. Such an operator is constrained by a wide variety of processes that can be measured experimentally. The strongest constraints are set from bounds on the electron electric dipole moment, flavor changing neutral currents and star cooling. Lab experiments on spin-dependent interactions provide weaker constraints but could plausibly be improved in the future.

The purpose of this paper was to explicitly derive the dipole operator in generic string theory orientifold compactifications at large volume and with the SM realized on intersecting stacks of D7-branes. We analyzed both brane photons and RR photons to find that the effect is IR dominated and the suppression scale of the operator is $\Lambda=\alpha\Ms$ with $\alpha$ a numerical factor. These results allow experimental and observational data on the dipole operator to be translated into quantitative bounds on the string scale. They extend from a conservative $1-100$~TeV in purely laboratory-based force measurements, via $10^2-10^5$~TeV in astrophysical observations, to a perhaps more optimistic $10^5-10^8$~TeV in flavor violating decays and electric dipole moments. The latter bounds assume significant CP and flavor violation in our operator as well as ${\mathcal{O}}(1)$ kinetic mixing. 

In anisotropic compactifications with a single large dimension, our bounds can exceed those derived from tests of 4d Newtonian gravity by several orders of magnitude. In scenarios with two or more extra dimensions, they exclude a substantially broader range of string scale values than the LHC constraints. This exclusion region, spanning multiple orders of magnitude, is however subject to independent constraints that arise if one insists on a reasonably explicit moduli stabilization framework, such as the LVS. Specifically, the requirement that SUSY is broken at a TeV scale or above imposes an upper limit on the compactification volume. The resulting bound on the string scale in the LVS is then comparable to or stronger than the present bound from the dipole operator. These bounds on the volume from the SUSY breaking scale may be somewhat relaxed if one considers a SUSY breaking sector which couples to the SM with maximal strength. In this case, the constraints we derived 
have direct relevance already at the current level of experimental sensitivity.

Thus, dipole interactions provide a unique way of probing a generic feature of string theory -- a photoverse of extra massless gauge bosons. Further phenomenological and experimental attention to them can hopefully increase their constraining power significantly.

Finally, the presence of SUSY at the fundamental level and its breaking offer a wealth of further opportunities: The superpartners of the massless hidden photons can give rise to a photinoverse, featuring mixing effects with the SM sector with observable signatures~\cite{Ibarra:2008kn, Arvanitaki:2009hb, Goodsell:2011wn}. The cosmological implications of this photinoverse, as well as of the photoverse discussed earlier, deserve further attention.

\subsection*{Acknowledgements}

We are very grateful for useful discussions with Björn Hassfeld and Jakob Moritz. This work was supported by Deutsche Forschungsgemeinschaft (DFG, German Research Foundation) under Germany’s Excellence Strategy EXC 2181/1 - 390900948 (the Heidelberg STRUCTURES Excellence Cluster). JS acknowledges support by the \textit{International Max Planck Research School for Precision Tests of Fundamental Symmetries}.


\appendix
\makeatletter
\DeclareRobustCommand{\@seccntformat}[1]{%
  \def\temp@@a{#1}%
  \def\temp@@b{section}%
  \ifx\temp@@a\temp@@b
  \appendixname\ \thesection:\quad%
  \else
  \csname the#1\endcsname\quad%
  \fi
} 
\makeatother
\renewcommand{\theequation}{A.\arabic{equation}}

\section{Conventions}
\label{app:conv}

\subsection{Indices and Hodge star}

In this appendix we shall enumerate the conventions employed in this paper. We mainly follow \cite{Martucci:2005rb} but similar conventions are also used in \cite{Myers:1999ps}.

As summarized in table~\ref{tab:indices}, we use capital letters of the latin alphabet for 10d spacetime indices, i.e. $A,B,\dots,M,N,\dots\,=0,\dots,9$ and greek letters from the beginning of the alphabet to denote curved spacetime directions along a D7-brane, i.e. $\alpha,\beta,\dots\,=0,\dots,7$. Lower case letters at the beginning of the latin alphabet are used for six-dimensional theories ($a,b,\dots\,=0,\dots,5)$ while letters in the middle of the greek alphabet $\mu,\nu,\dots\,=0,\dots,3$ are used for the standard four dimensions. We differentiate between curved and flat spacetime indices by underlining flat ones, i.e.
\begin{equation}
G_{MN}=e_{M}\,^{\underline A}e_{N}\,^{\underline B}\eta_{\underline A\underline B}\,, 
\end{equation}
and $\eta=\text{diag}(-,+,+,\dots,+)$.
We chose the Levi--Civita symbol in $d$ dimensions to take values $\pm1$ with all curved spacetime indices to be upstairs: $\varepsilon^{M_1\cdots M_{10}}=\pm1$. This means that the Levi--Civita symbol transforms as a tensor density. In $d$ spacetime dimensions and for a $p$-form $\omega$, we define the Hodge star operator like
\begin{equation}
(\star\,\omega)_{M_{p+1}\cdots M_d}=-\frac1{p!\sqrt{-g}}\omega^{M_1\cdots M_p}\varepsilon_{M_1\cdots M_d}\,.
\end{equation}
With this definition we have
\begin{equation}
\int\omega\wedge\star\,\omega=\frac1{p!}\int\dd^d\xi\sqrt{-g}|\omega|^2\,.   
\end{equation}
Also, as always, $\star\star\omega=(-1)^{p(d-p)+1}\omega$. In our conventions (in particular when $\Gamma_{(10)}\psi_M=+\psi_M$) the famous self-duality of $F_{(5)}$ is written as
\begin{equation}
-F_{(5)}=\star F_{(5)}\,.
\end{equation}
In Euclidean signature (e.g. for the compact CY space) we take 
\begin{equation}
(\star\,\omega)_{m_{p+1}\cdots m_d}=\frac1{p!\sqrt{g}}\omega^{m_1\cdots m_p}\varepsilon_{m_1\cdots m_d}\,.
\end{equation}
\subsection{Gamma matrices and fermions}

We write the $32\times32$ 10d flat gamma matrices as $\hat\Gamma^{\underline A}$; the 8d $16\times16$ gamma matrices are denoted by $\Gamma^{\underline\alpha}$; the 6d ones by $\tilde\gamma^{\underline a}$ and as usual the 4d ones by $\gamma^{\underline \mu}\,$. Further, we choose
\begin{equation}
\{\hat\Gamma^{\underline A},\hat\Gamma^{\underline B}\}=2\eta^{\underline A\underline B}\,,\qquad(\hat\Gamma^{\underline A})^\dagger=\hat\Gamma^{\underline 0}\hat\Gamma^{\underline A}\hat\Gamma^{\underline0}\,,
\end{equation}
and analogous relations in $8,\,6$ and $4$ dimensions. We define the chirality matrices as follows,
\begin{equation}
\hat\Gamma_{(10)}=\hat\Gamma^{\underline0\cdots\underline9}\,,\quad\Gamma_{(8)}=i\Gamma^{\underline 0\cdots\underline7}\,,\quad\tilde\gamma_{(6)}=\tilde\gamma^{\underline0\cdots\underline5}\,,\quad\gamma_{(4)}=i\gamma^{\underline 0\cdots\underline 3}\,.
\end{equation}
For an arbitrary chirality matrix $\Gamma$, the corresponding projector is given by $P_\pm\equiv(\one\pm\Gamma)/2$.

In $d$ Lorentzian dimensions we have the following identity,
\begin{equation}
\Gamma^{M_1\cdots M_p}\Gamma_{(d)}=s\frac{(-1)^{\lfloor p/2\rfloor+1}}{\sqrt{-g}(d-p)!}\varepsilon^{M_1\dots M_d}\Gamma_{M_{p+1}\cdots M_d}\,,\label{eq:gamma5tostar}
\end{equation}
where 
\begin{equation}
\Gamma_{(d)}=s\Gamma^{\underline 0\cdots\underline{D-1}}\,,\quad s\in\{\pm 1,\pm i\}\quad\text{s.t.}\quad\Gamma_{(d)}^2=\one\,.
\end{equation}

For a fermion field $\Psi$ we define the Dirac bar $\overline\Psi=i\Psi^\dagger\Gamma^{\underline 0}$ and for two Grassmann objects, conjugation also exchanges position, i.e. $(\xi\chi)^*=\chi^*\xi^*$.

Finally, $\sigma_i$ denotes are the usual Pauli matrices:
\be
\sigma_1=\begin{pmatrix}0&1\\1&0\end{pmatrix},\qquad\sigma_2=\begin{pmatrix}0&-i\\i&0\end{pmatrix}\quad\text{and}\quad\sigma_3=\begin{pmatrix}1&0\\0&-1\end{pmatrix}.
\ee

\section{Explicit reduction calculations}\label{app:Red}
\renewcommand{\theequation}{B.\arabic{equation}}

\subsection{Reduction to D7-branes intersection}

Here, we mainly follow \cite{Marchesano:2010bs} and our starting point is \eqref{eq:S8SYM2}. The holomorphic coordinate of $\mathbb C_2$ is denoted by $v=x^6+ix^7$ as in section \ref{subsec:RedTo6D} and the 6d coordinates are collectively denoted by $x$. Choosing a decomposition
\begin{equation}
\vartheta=\begin{pmatrix}\Psi_+(x)\otimes \eta_+&\Psi_+(x)\otimes \eta_-&\Psi_-(x)\otimes \eta_+&\Psi_-(x)\otimes \eta_-\end{pmatrix}
\begin{pmatrix}\zeta_{++}(v,\bar v)\\\zeta_{+-}(v,\bar v)\\\zeta_{-+}(v,\bar v)\\\zeta_{--}(v,\bar v)\end{pmatrix},\label{eq:basis6}
\end{equation}
the operator $\tildeslashed{D}$ has a particularly neat block structure (remember that we set $q=1$),
\begin{equation}
\tilde{\slashed D}=\begin{pmatrix}0&\mathbb D\\ \mathbb D^\prime&0\end{pmatrix}\quad\text{ with }\quad\mathbb D=\begin{pmatrix}i\overline\varphi&-2\partial\\-2\overline\partial&-i\varphi\end{pmatrix},\quad\mathbb D^\prime=\begin{pmatrix}-i\varphi&2\partial\\2\overline\partial&i\overline\varphi\end{pmatrix}. \label{eq:DD}
\end{equation}

Taking $\vev\varphi=vm\mathbf H$ as in \eqref{eq:phiVEV} with $m\sim1$ and real positive, i.e. large intersection angles between the intersecting branes, one expects a massless zero mode and a string scale ``intersection tower'' (see \cite{Marchesano:2010bs} for the full tower and corrections due to large angles). In what follows we shall only be concerned by the massless mode which, from \eqref{eq:DD}, has to satisfy 
\begin{equation}
\mathbb D\begin{pmatrix}\zeta_{-+}(v,\bar v)\\\zeta_{--}(v,\bar v)\end{pmatrix}=0\quad\text{ and }\quad\mathbb D^\prime\begin{pmatrix}\zeta_{++}(v,\bar v)\\\zeta_{+-}(v,\bar v)\end{pmatrix}=0\,.\label{eq:0ModInt}
\end{equation}
A straightforward computation shows that the only normalizable solution to \eqref{eq:0ModInt} is then
\begin{equation}
\vartheta_0=\Psi^0_-(x)\otimes\big(\eta_+-i\eta_-\big)\mathcal N\exp(- m|v|^2/2)\,,\label{eq:0Mode_app}
\end{equation}
where $\mathcal N$ is a normalization factor.

We are now able to derive the couplings of the intersection modes to the SUGRA background. Using \eqref{eq:Gamma8to6}, the couplings in \eqref{eq:S8flux} can be decomposed into a product of bilinears in 6d and internal bilinears involving $\zeta_0\equiv \mathcal N\exp(-m|v|^2/2)\begin{pmatrix}0 & 0 & 1 & -i\end{pmatrix}^T$ as follows,
\begin{align}
\overline\vartheta_0\Gamma^{\alpha\beta\gamma}\vartheta_0 \mathcal A_{\alpha\beta\gamma}=&\,\overline\Psi_-^0\tilde\gamma^{abc}\Psi_-^0 \mathcal A_{abc}(\zeta^\dagger_0\zeta_0)+3\overline\Psi_-^0\tilde\gamma^{ab}\tilde\gamma_{(6)}\Psi_-^0 \mathcal A_{ab \underline k}\big(\zeta^\dagger_0[\sigma_k]\zeta_0\big)\nonumber\\
&+6i\overline\Psi_-^0\tilde\gamma^{a}\Psi_-^0 \mathcal A_{a\underline{67}}\big(\zeta^\dagger_0[\sigma_3]\zeta_0\big)\,,\nonumber\\[5pt]
\overline\vartheta_0\Gamma^{\alpha\beta}\vartheta_0 \mathcal E_{\alpha\beta}=&\,\overline\Psi_-^0\tilde\gamma^{ab}\Psi_-^0\mathcal E_{ab}(\zeta^\dagger_0\zeta_0)+2\overline\Psi_-^0\tilde\gamma^{a}\tilde\gamma_{(6)}\Psi_-^0\mathcal E_{a \underline k}\big(\zeta^\dagger_0[\sigma_k]\zeta_0\big)\\
&+2i\overline\Psi_-^0\Psi_-^0\mathcal E_{\underline{67}}\big(\zeta^\dagger_0[\sigma_3]\zeta_0\big)\,,\nonumber\\[5pt]
\overline\vartheta_0\Gamma^{\alpha\beta}\Gamma_{(8)}\vartheta_0\mathcal U_{\alpha\beta}=&\,\overline\Psi_-^0\tilde\gamma^{ab}\tilde\gamma_{(6)}\Psi_-^0\mathcal U_{ab}(\zeta^\dagger_0[\sigma_3]\zeta_0)+2\overline\Psi_-^0\tilde\gamma^{a}\Psi_-^0\mathcal U_{a\underline k}(\zeta^\dagger_0[\sigma_k\sigma_3]\zeta_0)\nonumber\\
&+2i\overline\Psi_-^0\tilde\gamma_{(6)}\Psi_-^0\mathcal U_{\underline {67}}(\zeta^\dagger_0\zeta_0)\,.\nonumber
\end{align}
Here, $\mathcal A_{\alpha\beta\gamma}$ stands for either $H_{(3)\alpha\beta\gamma}$, $(\star_8 F_{(5)})_{\alpha\beta\gamma}$ or $F_{(5)\alpha\beta\gamma\underline{89}}$ and $\mathcal E_{\alpha\beta}$ and $\mathcal U_{\alpha\beta}$ are given by
\begin{align}
\mathcal E_{\alpha\beta}&\equiv \frac{1}{16}\left[H_{(3)\alpha\beta u}+H_{(3)\alpha\beta\bar u}-\frac{ie^\phi}{8}\left(F_{(3)\alpha\beta u}-F_{(3)\alpha\beta\bar u}\right)\right]\,,\nonumber\\
\mathcal U_{\alpha\beta}&\equiv \frac{1}{16}\left[H_{(3)\alpha\beta u}-H_{(3)\alpha\beta\bar u}-\frac{ie^\phi}{8}\left(F_{(3)\alpha\beta u} +F_{(3)\alpha\beta\bar u}\right)\right]\,.
\end{align}
The matrices $[\sigma_i]$ (with $\sigma_5\equiv\sigma_1$ and $\sigma_6\equiv\sigma_1$ in the expressions above for notational convenience), act on the internal spinor and can be obtained by taking the original Pauli matrices and going to the basis defined in \eqref{eq:basis6}. This results in 
\begin{equation}
[\sigma_i]=\begin{pmatrix}&\sigma_i\\\sigma_i&\end{pmatrix}.
\end{equation}

If the background varies slowly across the intersection (i.e. $\digamma(x^0,\dots,x^7,x^8,x^9)\approx \digamma(x^0,\dots,x^7,0,0)$ for any background $\digamma$) and the sum in \eqref{eq:basis6} is restricted to the zero mode \eqref{eq:0Mode_app}, the coupling terms can just be read off\footnote{One more term from $\mathcal U_{\alpha\beta}$ survives but it cannot have two 4d indices as we require for our dipole coupling of interest, so we do not write it here. Also, the terms involving two gamma matrices vanish due to the 6d chirality of the zero mode.},
\begin{align}
\label{eq:S6Dflux}
S^{\rm flux}_{\rm 6d,0}=\mu\int \dd^6\xi\sqrt{-g}\tr\Bigg\{&\frac1{12}\overline\Psi^0_-\tilde\gamma^{abc}\Psi^0_-\left[H_{(3)abc}+\frac\ep4 \big(F_{(5)abc\underline{89}}-(\star_8 F_{(5)})_{abc}\big)\right]\Bigg\}\,.
\end{align}
In this action the normalization $\mathcal N$ as well as factors arising from the integral over $\mathbb C_2$ have been absorbed in $\mu$.

\subsection{Estimate of RR photon prefactors}
\label{app:RRTor}

We consider a rectangular flat torus with coordinates $y^1,\dots,y^6$ ranging from zero to one and of radius $R$. The metric is given by 
\begin{equation}
{\rm d}s^2=(2\pi R)^2{\rm d}y^i{\rm d}y^i\,.
\end{equation}
Since the torus is rectangular, a possible choice of vielbein is ${e_i}^{\underline i}=(2\pi R)\delta_i^{\underline i}$. Further we assume two D7s wrapping the directions $1234$ and $1256$ respectively. The three-forms $\alpha=\dd y^1\wedge\dd y^4\wedge\dd y^5$ and $\beta=\dd y^2\wedge \dd y^3\wedge\dd y^6$ obey
\begin{equation}
\star_6\,\alpha=\beta \qquad\mbox{and} \qquad \int_{T^6} \alpha\wedge\beta=1\,.
\end{equation}

While $\{\alpha,\beta\}$ can be extended to a full symplectic basis in the sense of \eqref{eq:symplectic} we will not do so and focus only one these two chosen three-forms. The matrices of \eqref{eq:abstar} are then one-by-one and read $A=D=0$ and $C=-B=-1$. Thus, defining the `effective $\underline{89}$-directions' democratically\footnote{
As explained in sect.~\ref{mat_curv_to_4d}, we do not treat the case of large-angle intersections carefully enough to be certain about ${\cal O}(1)$ numbers potentially arising from this choice.
}
as $\tilde e_{\underline 8}=(e_{\underline 3}+ e_{\underline 5})/\sqrt 2$ and $\tilde e_{\underline 9}=(e_{\underline 4}+e_{\underline 6})/\sqrt 2$, equation \eqref{eq:fRR} can be evaluated explicitly,
\begin{equation}
f(z)={ \sqrt{2}\kappa}\left(\alpha_{\underline{z89}}-i\beta_{\underline{z89}}\right)=-\frac{ \sqrt{2}\kappa}{16\pi^3R^3}\,.
\end{equation}
The point of this exercise was merely to convince ourselves that (up to a factor 1/2, which is anyway not reliable given the simplistic nature of our toy model) we find a suppression by the volume $(2\pi R)^3$ of the relevant cycle. In particular, in contrast to the brane-photon case, there are no high powers of $(2\pi)$ or similar numerical surprises.

\subsection{Magnetized torus}
\label{app:MagTor}

In this appendix we focus on the reduction from 6d to 4d in the case where the intersection cycle is a flat two-torus with Teichmüller parameter $\tau$, which allows for explicit computations. We assume a trivial metric, such that $\tau$ specifies the identifications on $\mathbb{C}$ that define the torus. We consider a magnetized torus by following \cite{Bachas:1995ik,Cremades:2004wa,Marchesano:2010bs,Buchmuller:2016gib}. Denoting the coordinates of the torus by $z=x^4+ix^5$, we have the flux
\begin{equation}
F_{(2)}=-f\,\dd x^4\wedge \dd x^5=\frac i2f\,\dd \overline z\wedge \dd z\,,\label{eq:FluxChoice}
\end{equation}
where the minus sign has been introduced for future convenience. Flux needs to be quantized, i.e. $f\, \im \tau/2\pi \in\mathbb Z$ (or supplemented by dimension-full factors depending on conventions). In what follows we restrict ourselves to $f>0$ and a possible choice of gauge potential is
\begin{equation}
A_{(1)}=\frac i4f\big(\overline z\,\dd z-z\,\dd\overline z\big)\,.\label{eq:FluxGaugPot_app}
\end{equation}

Once the flux is turned on, translation invariance is broken and all fields are only periodic up to a gauge transformation. In particular, we have 
\begin{equation}
A_{(1)}(z+1)=A_{(1)}(z)+\dd\Lambda_1\quad\text{ and }\quad A_{(1)}(z+\tau)=A_{(1)}(z)+\dd\Lambda_\tau\,,
\end{equation}
where the choice of $\Lambda_{1,\tau}$ compatible with \eqref{eq:FluxGaugPot_app} is given by
\begin{equation}
\Lambda_1=\frac i4f\big(z-\overline z\big)\quad\text{ and }\quad\Lambda_\tau=\frac i4f\big(\overline\tau z-\tau\overline z\big)\,.
\end{equation}
This also implies that any field charged under the corresponding $U(1)$ with a charge $q$ (in the case of interest $U(1)_{\mathbf H}$) will also be quasiperiodic,
\begin{equation}
\Psi(z+1)=\exp\big(iq\Lambda_1)\psi(z)\quad\text{ and }\quad\Psi(z+\tau)=\exp\big(iq\Lambda_\tau)\psi(z)\,.\label{eq:BouQuasiPer}
\end{equation}
In the following we choose to normalize $\mathbf H$ such that $q\equiv1$.

The 6d zero mode $\Psi_-^0$ can be decomposed as follows:
\begin{align}
\label{eq:decomp_KK}
\Psi^0_-(x,z,\overline z)&=\sum_n\left(\psi_{+n}\otimes \eta_-+\psi_{-n}\otimes \eta_+\right)\\
&\equiv\sum_n\left(\psi_{+n}(x)\otimes \eta_-\quad\psi_{-n}(x)\otimes \eta_+\right)\begin{pmatrix}\zeta_{+-n}(z,\overline z)\\ \zeta_{-+n}(z,\overline z)\end{pmatrix}\,.
\end{align}
The sign in the subscript indicates the chirality according $SO(3,1)\times SO(2)$, i.e. $\gamma_{(4)}\psi_\pm=\pm\psi_\pm$ and $\eta_+=(1\ 0)^T$, $\eta_-=(0\ 1)^T$. The index $n$ labels the KK levels so that the $\psi_{\pm n}$'s have the same masses,  and the 4d cordinates are collectively denoted by $x$. The 6d kinetic term then splits as,
\begin{equation}
    S_{\rm 6d,0}^{\rm YM}\supset\mu\int_{\mathcal D}\dd^6\xi\sqrt{-g}\,\tr\,\bigg\{\overline\Psi_-^0 \tilde\gamma^\mu\nabla_\mu\Psi_-^0+
    2\sum_n\begin{pmatrix}\overline\psi_{-n}&\overline\psi_{+n}\end{pmatrix}
    \underbrace{\begin{pmatrix} 0 & \nabla_z\\
-\nabla_{\bar z} & 0 \end{pmatrix}}_{\textstyle  \equiv\,\,\hatslashed{D}}
\begin{pmatrix}\psi_{-n}\\\psi_{+n}\end{pmatrix}\bigg\}\,,\label{eq:S6SYM2}
\end{equation}
and we are interested in the full tower of eigenmodes of the operator $\hatslashed{D}=\tilde\gamma^{a'}\nabla_{a'}$. Along the torus coordinates the covariant derivative is $\nabla_z=\partial_z-iA_{(1)z}$. Note that the presence of the minus sign in front of $\nabla_{\overline z}$ originates from the $\gamma_{(4)}$ involved in the definition of $\tilde\gamma^{a^\prime}$. Further, $\hatslashed{D}$ should also include a dimensionful factor of inverse length due to the vielbein present in the $\tilde\gamma^{a'}\nabla_{a'}$ contraction. We shall omit it in the sequel to improve readability.

From the definition of the field strength in terms of covariant derivatives, it is immediately clear that
\begin{equation}
[\nabla_z,\nabla_{\overline z}]=-iF_{z\overline z}=-\frac12f\,,
\end{equation}
where for the last equality, we have used the particular choice \eqref{eq:FluxChoice}. Thus, by defining $a=-i\nabla_{z}/\sqrt {f/2}$ and $\overline a=-i\nabla_{\overline z}/\sqrt {f/2}$, one finds operators that fulfill the usual harmonic oscillator algebra,
\begin{equation}
[a,\overline a]=1\,.
\end{equation}
The Dirac operator then reads
\begin{equation}
\tilde\gamma^{a^\prime}\nabla_{a^\prime}=2\sqrt {f/2}\begin{pmatrix}&i\,a\\-i\,\overline a&\end{pmatrix},\label{eq:Dirac}
\end{equation}
and the Klein--Gordon operator is
\begin{equation}
\big(\tilde\gamma^{a^\prime}\nabla_{a^\prime}\big)^2=2f\begin{pmatrix}\overline aa+1&\\&\overline aa\end{pmatrix}.\label{eq:KleinGordon}
\end{equation}

Hence, it remains to find a function $\zeta_0(z,\overline z)$ obeying the boundary conditions of \eqref{eq:BouQuasiPer} and that is annihilated by $a$. Then the full tower of eigenmodes can be expanded like
\begin{equation}
\Psi^0_-=\sum_i\big(\psi_{+i}(x)\otimes\eta_-+\psi_{-i}(x)\otimes\eta_+\big)\frac{\overline a^i}{\sqrt{i!}}\zeta_0(z,\overline z)\,,
\end{equation}
where we decompose the field into modes indexed by $i$ which share the same internal profile. This index is not the KK level as in \eqref{eq:decomp_KK}. Indeed, from \eqref{eq:KleinGordon} one can see that $\psi_{+i}$ has a KK mass given by $\sqrt{2fi}$ while $\psi_{-i}$ has mass $\sqrt{2f(i+1)}$ and the Dirac operator \eqref{eq:Dirac} couples $\psi_{+i}$ to $\psi_{-(i-1)}$ with a mass term.

The number of ground-state functions depends on the flux integer $N\equiv\im \tau f/2\pi$,
\begin{equation}
\zeta^p_0(z,\overline z)=\exp\Big(-\frac f4\overline z( z- \overline z)\Big)\theta\jac(p/N,0)(N\overline z,-N\overline\tau)\,,\quad p=0,\dots,N-1\,,
\end{equation}
where $\theta$ is the Jacobi theta function,
\begin{equation}
\theta\jac(a,b)(\nu,\tau)=\sum_{n\,\in\,\mathbb Z}\exp\big(\pi i(a+n)^2\tau\big)\exp\big(2\pi i(a+n)(\nu+b)\big)\,.
\label{eq:Jacobi}
\end{equation}

In particular, this implies that
\begin{align}
\int \dd z\,\dd \overline z\,\overline\Psi^0_-\hatslashed{D}\Psi^0_-=&2\int\dd z\,\dd\overline z\,\sqrt{f/2}\begin{pmatrix}\overline\psi^p_{-i}(\zeta^p_i)^*&\overline\psi^p_{+i}(\zeta^p_i)^*\end{pmatrix}
\begin{pmatrix}
    &i\, a\\ -i\,\bar a&
\end{pmatrix}
\begin{pmatrix}\psi^q_{-j}\zeta^q_j\\\psi^q_{+j}\zeta^q_j\end{pmatrix}\nonumber\\
=&-i\sqrt{2(j+1)f}\,\overline\psi^p_{+(j+1)}\psi^p_{-j}+\text{h.c.}\label{eq:4DKinRed}
\end{align}
Coming back to the KK decomposition, the internal profiles $\zeta_{+-n}$ and $\zeta_{-+n}$ are given in table \ref{tab:Sigma}. Famously, the zero mode yields a chiral 4d spinor, and the internal profiles are the same for the left- and right-handed components, simply shifted by one level. The background couplings of \eqref{eq:S6SUGRA} with strictly two 4d spacetime indices $\mu,\nu$ then reduce to
\begin{multline}
    \begin{pmatrix}\overline\psi^p_{-n}(\zeta^p_{n})^*&\overline\psi^p_{+n}(\zeta^p_{n-1})^*\end{pmatrix}
\begin{pmatrix}
    0 & \mathcal A_{\mu\nu z}(z,\bar z)\\ -\mathcal A_{\mu\nu \overline z}(z,\bar z) & 0
\end{pmatrix}
\begin{pmatrix}\gamma^{\mu\nu}\psi^q_{-m}\zeta^q_{m}\\\gamma^{\mu\nu}\psi^q_{+m}\zeta^q_{m-1}\end{pmatrix}=\\[5pt]
=\overline\psi^p_{-n}\gamma^{\mu\nu}\psi^q_{+m}\left[\mathcal A_{\mu\nu z}(\zeta^{p}_{n})^*\zeta^q_{m-1}\right]-\overline\psi^p_{+n}\gamma^{\mu\nu}\psi^q_{-m}\big[\mathcal A_{\mu\nu \overline z}(\zeta^{p}_{n-1})^*\zeta^q_{m}\big]\,,\label{eq:4DHcoup}
\end{multline}
where $\mathcal A_{\mu\nu z}$ stands for either $H_{(3)\mu\nu z}$, $(\star_8 F_{(5)})_{\mu\nu z}$ or $F_{(5)\mu\nu z\underline{89}}$ (and similarly for $\mathcal{A}_{\mu\nu\bar z}$). Note that the two terms in the second line are precisely conjugate to each-other, since $(\overline \psi\gamma^{\mu\nu}\chi)^*=-\overline\chi\gamma^{\mu\nu}\psi$. From \eqref{eq:4DHcoup}, it is clear that the chiral massless mode does not couple to itself. However, it is also clear that, if $\mathcal A\neq0$, it induces a coupling between the zero mode and states of the massive Kaluza--Klein tower.

\begin{table}
    \centering
    \begin{tabular}{c|cc}
       KK level $n$  & $\zeta_{+-n}(z,\bar z)$& $\zeta_{-+n}(z,\bar z)$ \\\hline\hline
        $0$ & $\zeta^p_0(z,\overline z)$ & $-$ \\
        $1$ & $\zeta^p_1(z,\overline z)$ & $\zeta^p_0(z,\overline z)$\\
        $2$ & $\zeta^p_2(z,\overline z)$ & $\zeta^p_1(z,\overline z)$\\
        $\vdots$ & $\vdots$&$\vdots$\\
    \end{tabular}
    \caption{The internal profiles of the KK modes. Note that the zero mode gives rise to chiral matter in 4d as expected.}
    \label{tab:Sigma}
\end{table}

As in the general case, this coupling between the zero mode and higher KK excitations together with the KK mass terms and the Yukawa coupling back to the zero mode, allows dipole interactions of the form \eqref{eq:resXcoup} through the tree-level process of figure \ref{fig:mediationTree}. In the explicit case of a flat torus and assuming smooth behavior for $\mathcal A$ one can show that $\sum_n|c_{(n)}|<\infty$, confirming IR domination and justifying our EFT approach \cite{Marchesano:2010bs}. 

\section{\bm Estimating Loop corrections to complex-structure $F$-terms}
\label{app:Fterm}
\renewcommand{\theequation}{C.\arabic{equation}}

The leading order Kähler potential,
\be
K_0=-\log(S+\overline S)-\log\Big(-i\int\Omega\wedge\overline\Omega\Big)-2\log(\mathcal V)\,,
\ee
generically acquires $\ap$ and $\gs$ corrections,
\be
K\,=\,K_0+\delta K \,\equiv\, K_0+\delta K_{(\ap)}+\delta K_{(\gs)}\,.\label{eq:K_Korr}
\ee
In particular, $\delta K_{(\ap)}$ is the famous BBHL \cite{Becker:2002nn} correction
\be
\delta K_{(\ap)}=-\frac{\xi\,\Re(S)^{3/2}}{\mathcal V}\,,
\ee
which plays a central role in the LVS. Here $S$ is the axio-dilaton and $\Re(S)=\gs^{-1}$. For loop corrections to the Kähler potential \cite{vonGersdorff:2005bf, Berg:2005ja, Cicoli:2007xp}, a universal expression has been proposed in \cite{Berg:2007wt}. A closely related but more general form has been argued in \cite{Gao:2022uop},
\be
\delta K_{(\gs)}=\delta K^{\rm KK}_{(\gs)}+\delta K^{\rm W}_{(\gs)}\sim-\frac{C^{\rm KK}(U_a,\overline U_a)\mathcal T(T_i)}{\Re(S)\vol}-\frac{C^W(U_a,\overline U_a)}{\mathcal I(T_i)\vol} \,.
\label{eq:loop_cor}
\ee
Here, $U_a$ are complex structure moduli, $T_i$ are the K\"ahler moduli, and $\mathcal T$, $\mathcal I$ are homogeneous functions of degree $1/2$ in $T_i$.\footnote{The notation $C^{\rm KK}$ and $C^{\rm W}$ refers to `Kaluza--Klein' and `winding', based on specific diagrams contributing in the torus-orientifold model studied in \cite{Berg:2007wt}. As argued in \cite{Gao:2022uop}, the names `local correction' and `genuine loop correction' may be more appropriate in the generic Calabi--Yau case.} As an illustration, we give a more explicit expression for the case of a single K\"ahler modulus,
\be
K=-3\log(T+\overline T)-\frac{\xi\,\Re(S)^{3/2}}{(T+\overline T)^{3/2}}-\frac{C^{\rm KK}(U_a,\overline U_a)}{\Re(S)(T+\overline T)}-\frac{C^{\rm W}(U_a,\overline U_a)}{(T+\overline T)^2}\,.
\ee

In an LVS-type setup, the $F$-terms $F^S$ and $F^{U_a}$ acquire non-zero contributions in at least five ways:
\begin{itemize}
\item The inverse of the corrected Kähler metric $K_0+\delta K$ mixes the $(S,U_a)$ with the $T_i$ directions. Given the non-zero value of $D_{T_i}W$ in the LVS AdS minimum, this induces non-zero $F$-terms for $S$ and $U_a$.
\item The correction $\delta K$ directly modifies the covariant derivatives $D_{T_i,S,U_a}$.
\item$W_{\rm np}$ depends on $U_a$ and $S$ via its prefactor: $W_{\rm np}=A(U_a,S)\,e^{-a_iT^i}$. 
\item The expectation values of $S$ and $U_a$ are shifted with respect to the leading-order `GKP-level' vacuum.
\item The uplift to dS generically induces non-zero $F$-terms.
\end{itemize}
Following \cite{Conlon:2005ki,Blumenhagen:2009gk,Aparicio:2014wxa} we disregard the last two effects, postponing their analysis to future work. In practice, this amounts to assuming that, even in the stabilized and uplifted vacuum, the leading-order relation $\partial_{S,U_a}W+(K_0)_{S,U_a}W=0$ remains approximately valid. The size of $F^{U_a}$ induced by $W_{\rm np}$ has been estimated in \cite{Aparicio:2014wxa}:
\be
\kappa_4F^{U_a}_{\rm np}\sim\frac{W_0}{\vol^2}\,.
\ee
We now proceed by checking that the contributions induced by $\delta K$ do not exceed this parametrically. Given our assumptions, one has
\begin{align}
\kappa_4F^{U_a}_{\text{K\"ahler}}=\,\,&e^{K/2}\big(K^{U_a\overline U_b}D_{\overline U_b}\overline W+K^{U_a\overline S}D_{\overline S}\overline W+K^{U_a\overline T_i}D_{\overline T_i}\overline W\big)\nonumber \\
    \approx\,\,&e^{K/2}\overline W_0\big(K^{U_a\overline U_b}\delta K_{\overline U_b}+K^{U_a\overline S}\delta K_{\overline S}+K^{U_a\overline T_i}K_{\overline T_i}\big)\,.
\end{align}
The inverse metric can generically be approximated by 
\be
K^{\overline A B}=K_0^{\overline A B}-K_0^{\overline A C}\delta K_{C\overline D}K_0^{\overline D B}+K_0^{\overline A C}\delta K_{C \overline D}K_0^{\overline D E}\delta K_{E\overline F}K_0^{\overline F B}+\mathcal O\big(\delta K^3\big)\,,
\ee
where the indices $A,B,\dots$ take values $S,U_a,T_i$. One can now systematically expand $F^{U_a}$ in $\delta K$. Assuming a correction $\delta K_{(n)}$ which is homogeneous of degree $n$ in the 4-cycle volumes, one finds 
\be
\kappa_4\,\delta F_{(1)}^{U_a}\approx e^{K/2}\overline W_0\,K_0^{U_a\overline U_b}(\delta K_{(n)})_{\overline U_b}(1+n)\,.
\ee
Here, the index `(1)' characterizes the leading, linear order. Thus, by definition, $\delta F_{(1)}^{U_a}$ is linear in $\delta K$ and we can treat the two terms in \eqref{eq:loop_cor} independently. The `KK effect' has $n=-1$ and does hence obviously not contribute due to the same type of cancellation that is responsible for the extended no-scale structure \cite{vonGersdorff:2005bf,Berg:2007wt,Cicoli:2007xp}. Focusing on the scaling properties only, one then finds
\be
 \kappa_4\,\delta F_{(1)}^{U_a}\,\,\sim\,\, e^{K/2}\,K_0^{U_a\overline U_b} \,\left(\delta K^{\rm W}_{(\gs)}\right)_{\overline U_b} 
 \,\sim\, {\cal V}^{-2}\cdot\mathcal I^{-1}.\label{eq:F1}
\ee

At quadratic order one has
\begin{eqnarray}
\kappa_4\,\delta F_{(2)}^{U_a}
& \approx & e^{K/2}\overline W_0K_0^{U_a\overline U_b}\bigg(\delta K_{\overline U_b U_c}K_0^{U_c\overline U_d}\left(\delta K^W_{(\gs)}\right)_{\overline U_d}
\label{eq:F2}
\\&& +\delta K_{\overline U_b S}K_0^{S\overline S}\left(\delta K_{(\ap)}\right)_{\overline S}-\delta K_{ \overline U_bT_i}K_0^{T_i\overline T_j}\widetilde{\delta K}_{\overline T_j}\bigg)\,,
\nonumber
\end{eqnarray}
where\footnote{
The 
prefactors are simply the degrees of homogeneity in 4-cycle-variables of each contribution.
}
\be
\widetilde{\delta K}\equiv
-\frac{3}{2}\delta K_{(\alpha')}
-\delta K^{\rm KK}_{(\gs)}
-2\delta K^{\rm W}_{(\gs)}\,.
\ee
Similarly to \eqref{eq:F1}, no-scale type cancellations have occurred, simplifying the first two terms in \eqref{eq:F2}.

Now we want to understand 
how $\delta F_{(2)}^{U_a}$ scales with ${\cal V}$. This is non-trivial since we know neither the scaling of the functions $\mathcal I$ and $\mathcal T$ with the overall volume nor the explicit Kähler potential. Analyzing just the scaling in term of all 4-cycle variables is insufficient for the large volume scenario, where some 4-cycles may have a volume much smaller than ${\cal V}^{2/3}$. With some assumptions, a result can nevertheless be obtained:

The r.h.~side of \eqref{eq:F2} contains, inside the bracket, a sum of three terms. Disregarding ${\cal O}(1)$ prefactors, the first term scales as $\vol^{-2}(\mathcal T+1/\mathcal I)/\mathcal I$. The second term scales as $\vol^{-2}\mathcal T$. Finally, the contribution of the third term is somewhat more involved. It is convenient to discuss its factors $\delta K_{\overline U_bT_i}$ and $K_0^{T_i\overline T_j}\delta K_{\overline T_j}$ separately: The first factor, $\delta K_{\overline U_bT_i}$, contributes as
\begin{equation}
\frac1\vol\left(\mathcal T^\prime -\frac{\mathcal T \vol^\prime}{\vol}\right)\qquad\text{or}\qquad-\frac1{\mathcal I\vol}\left(\frac{\mathcal I^\prime}{\mathcal I}+\frac{\vol^\prime}\vol\right)\label{eq:F2T_terms}
\end{equation}
depending on whether one considers the KK or the winding-mode correction. Here the prime denotes taking the derivative with respect to $T_i$. The second factor has been estimated in \cite{Cicoli:2007xp} to scale as $K_0^{T_i\overline T_j}\delta K_{\overline T_j}\lesssim\vol^0\sim1$, where
the conjecture of \cite{Berg:2007wt} for the KK corrections has been used. While, on the one hand, this form was argued in \cite{Gao:2022uop} to be too restrictive, we are on the other hand not aware of an explicit KK-type correction violating the estimate of \cite{Cicoli:2007xp}. 

Thus, in total, the largest contribution to $\delta F_{(2)}^{U_a}$ arises from a combination of the prefactor $\exp(K/2)\sim 1/{\cal V}$, a factor $\sim 1/{\cal V}$ from \eqref{eq:F2T_terms}, and the factor $\sim {\cal V}^0$ just discussed. We note that a further assumption went into this result: We ascribed a scaling $\lesssim 1/{\cal V}$ to \eqref{eq:F2T_terms}. This assumption is natural since the additional factors involving $\mathcal T$ or $\mathcal I$ have negative scaling in terms of 4-cycle-variables. However, we can not rule out an enhancement of type `large cycle divided by small cycle'. The best we can say is that, specifically for LVS-type geometries, the analysis of \cite{Gao:2022uop} does not see such effects, independently of the conjecture of \cite{Berg:2007wt}. By the same logic, we assume that no volume enhancement comes from the factor $1/\mathcal I$ in \eqref{eq:F1}. As a result, our present best estimate for the volume scaling of complex-structure $F$-terms is
\begin{equation}
\kappa_4F^{U_a}=\kappa_4F^{U_a}_{(1)}+\kappa_4F^{U_a}_{(2)}\,\sim\,\frac{1}{\vol^2}\,.
\end{equation}
\vspace*{-1.8cm}

\bibliographystyle{JHEP}
\bibliography{references}

\end{document}